\documentclass{article}

\mathcode`:="003A
\usepackage{multirow}
\usepackage{amssymb}
\usepackage{amsthm}

\usepackage{diagrams}

\usepackage[a4paper, portrait, margin=2cm]{geometry}

\usepackage{MnSymbol}
\usepackage{stmaryrd}

\usepackage{tikz}
\usetikzlibrary{arrows}
\usetikzlibrary{fadings}

\usepackage{tikz-cd}
\tikzcdset{every label/.append style = {font = \small}}

\usepackage{varioref}

\newcommand{\defeq}{\stackrel{\vartriangle}{=}}

\newcommand{\defiff}{\stackrel{\vartriangle}{\iff}}

\newcommand{\Real}{\mathbb{R}}
\newcommand{\Nat}{\mathbb{N}}
\newcommand{\State}{\mathbb{S}}

\newcommand{\den}[1]{\ensuremath{\left\llbracket #1 \right\rrbracket}} 

\newcommand{\inv}[1]{{#1}^{-1}}

\newcommand{\X}{\times}

\newcommand{\absn}[1]{\lvert\thinspace {#1} \thinspace\rvert}

\newcommand{\eg}{e.\,g.}
\newcommand{\ie}{i.\,e.}

\newcommand{\interiorOf}[1]{\ensuremath{{#1}^{\circ}}}

\newcommand{\norm}[1]{\ensuremath{{\lVert\thinspace {#1}
\thinspace\rVert}}} 

\newcommand{\set}[1]{\ensuremath{\left\{{#1}\right\}}}
\newcommand{\setbarNormal}[2]{\ensuremath{\{{#1} \mid
{#2}\}}}

\newcommand{\PS}{\mathsf{P}}
\newcommand{\CS}{\mathsf{C}}
\newcommand{\OS}{\mathsf{O}}

\newcommand{\pL}{\mathbb{P}}
\newcommand{\cL}{\mathbb{C}}

\newcommand{\arrayoptions}[2]{\setlength{\arraycolsep}{#1}\renewcommand{\arraystretch}{#2}}

\newcommand{\Set}{\mathbf{Set}}
\newcommand{\Po}{\mathbf{Po}}

\newcommand{\Top}{\mathbf{Top}}
\newcommand{\C}{\mathbb{A}}
\newcommand{\id}{\text{id}}

\newcommand{\cl}[1]{\overline{#1}}

\newcommand{\ssp}[1]{\ell_{#1}}
\newcommand{\kssp}[1]{\cl{\ell}_{#1}}
\newcommand{\cub}[1]{B_{#1}}
\newcommand{\kcub}[1]{\cl{B}_{#1}}
\newcommand{\ol}[1]{\overline{#1}}
\newcommand{\ul}[1]{\underline{#1}}

\newtheorem{definition}{Definition}[section]
\newtheorem{proposition}[definition]{Proposition}
\newtheorem{lemma}[definition]{Lemma}
\newtheorem{corollary}[definition]{Corollary}
\newtheorem{theorem}[definition]{Theorem}

\newtheorem{remark}[definition]{Remark}
\newtheorem{example}[definition]{Example}

\newcommand{\Haus}{\mathbf{Haus}}
\newcommand{\KH}{\mathbf{KH}}
\newcommand{\MS}{\mathbf{Met}}
\newcommand{\CMS}{\mathbf{CMS}}
\newcommand{\KMS}{\mathbf{KMS}}
\newcommand{\NVS}{\mathbf{NVS}}
\newcommand{\BS}{\mathbf{Ban}}
\newcommand{\CL}{\mathbf{CL}}
\newcommand{\wCL}{\omega\CL}

\usepackage{float}
\floatstyle{ruled} 
\restylefloat{figure}
\restylefloat{table}

\title{Robustness, Scott Continuity, and Computability} \author{Amin Farjudian\thanks{School of Computer Science, University of
    Nottingham Ningbo China,
    Amin.Farjudian@gmail.com}
  \and Eugenio Moggi\thanks{DIBRIS, Genova Univ., Genova, Italy, moggi@unige.it}}

\date{}

\begin{document}

\maketitle

\begin{abstract}
  Robustness is a property of system analyses, namely monotonic maps
  from the complete lattice of subsets of a (system's state) space to
  the two-point lattice.  The definition of robustness requires the
  space to be a metric space.  Robust analyses cannot discriminate
  between a subset of the metric space and its closure, therefore one
  can restrict to the complete lattice of closed subsets. When the
  metric space is compact, the complete lattice of closed subsets
  ordered by reverse inclusion is $\omega$-continuous and robust
  analyses are exactly the Scott continuous maps. Thus, one can also
  ask whether a robust analysis is computable (with respect to a
  countable base). The main result of this paper establishes a
  relation between robustness and Scott continuity, when the metric
  space is not compact.  The key idea is to replace the metric space
  with a compact Hausdorff space, and relate robustness and Scott
  continuity by an adjunction between the complete lattice of closed
  subsets of the metric space and the $\omega$-continuous lattice of
  closed subsets of the compact Hausdorff space. We demonstrate the
  applicability of this result with several examples involving Banach
  spaces.
\end{abstract}



\tableofcontents

\section{Introduction}

The main contribution of this paper is relating \emph{robust analyses}
and \emph{Scott continuous maps} (between $\omega$-continuous
lattices).
This contribution is relevant to the broader endeavor of developing
software tools for system analysis based on mathematical models.
Typically, the behavior of a controlled system is given a priori, but
for most systems the open system approach is insufficient as the
correctness of the controlling system depends on properties of the
environment. This requires to model also the environment.
Software tools (for system analysis) manipulate formal descriptions.
The key point of formal descriptions is their mathematical exactness.
However, exactness should not be confused with precision.  In
particular, mathematical descriptions should make explicit \emph{known
  unknowns} and the amount of imprecision. There are two unavoidable
sources of imprecision: errors in measurements (on physical systems)
and representations of continuous quantities in software tools.

The key feature of robust analyses is the ability to cope with small
amounts of imprecision.
On the other hand, analyses can be implemented in software tools only
if they are \emph{computable}.  A definition of computability for
\emph{effectively given domains} has been proposed in
\cite[Definition~3.1]{smyth1977effectively}, where the domains
considered include those of interest for us, namely
$\omega$-continuous lattices. The key point of this, and similar
proposals, is that computable maps (between effectively given domains)
are necessarily Scott continuous.

For the benefit of readers, before outlining the main result of the
paper, we review the context in which it is placed and recall related
results, while keeping technicalities to a minimum.

\paragraph{Systems.}
First, one has to decide how to model systems.
The simplest systems, \ie, \emph{discrete} systems, can be modeled by
a set $\State$ (of states) and a transition map $t:\State \to \State$
describing the \emph{deterministic} state change of the system in one
step.
The systems we consider are \emph{closed}, \ie, they do not interact
with the \emph{environment}, or to put it differently, a model should
account also for the environment.
In this respect, it is important to model also \emph{known unknowns}
(for discrete systems, imprecision is not an issue). The simplest way
to do this is by \emph{non-determinism}, namely, a state in $\State$
is replaced by a set of states and the transition map is replaced by a
transition relation $T\subseteq\State\X\State$. In theoretical
computer science, the pair $(\State,T)$ is called a \emph{transition
  system}.

\paragraph{Analyses.}
We move from the category of sets and relations to the category of
complete lattices and monotonic maps. More precisely, we replace
$\State$ with the complete lattice $\pL(\State)$ of subsets of
$\State$, and a relation $T\subseteq\State\X\State'$ with the
monotonic map $T_*:\pL(\State) \to \pL(\State')$ such that:
\begin{equation*}
T_*(S)\defeq\{s' \mid \exists s:S.T(s,s')\}.
\end{equation*}

We take as partial order $\leq$ on $\pL(\State)$ reverse inclusion,
\ie, $S'\leq S\iff S'\supseteq S$.  The rational for this choice is
that a smaller set (of states) provides more information on the actual
state of the system. The transition relations on $\State$ form a
complete lattice $\pL(\State\X\State)$, where $T'\leq T$ means that
$T$ is more deterministic than $T'$.

Several \emph{analyses} correspond to monotonic maps between complete
lattices. For instance, reachability analysis for transition systems
on $\State$ corresponds to the map
$R:\pL(\State\X\State)\X\pL(\State)\to \pL(\State)$ given by
$R(T,I)\defeq T^*(I)$, \ie, the set of states reachable in a finite
number of steps from (a state in) $I$, while safety analysis
corresponds to the map
$S:\pL(\State\X\State)\X\pL(\State)\X\pL(\State)\to \Sigma$ in which
$\Sigma$ is the two-point lattice $\bot<\top$ and
$S(T,I,E)=\top\defiff T^*(I)\cap E=\emptyset$, \ie, no bad state in
$E$ is reachable from $I$.

\paragraph{Approximation.}
The partial order on a complete lattice $X$
allows (qualitative) comparisons, in particular we say that $x'$ is an
\emph{over-approximation} of $x$ when $x'\leq x$.
The category of complete lattices and monotonic maps is also the
natural setting for \emph{abstract interpretation}
\cite{cousot1996abstract,cousot1977abstract}.  More precisely, given
an interpretation $\den{-}$ of a (programming) language in a complete
lattice $X$, one can choose another complete lattice $X_a$ related to
$X$ by an \emph{adjunction} (also known as Galois connection)%
\begin{tikzcd}[column sep = large]
  X_a \arrow[r, "\top", "\gamma"', yshift = -1.1ex] & X \arrow[l,
  "\alpha"', yshift = 1.1ex],
\end{tikzcd}\ie, a pair of monotonic maps $\alpha$ (called abstraction) and
$\gamma$ (called concretization) such that
$x'\leq_a\alpha(x)\iff\gamma(x')\leq x$.
In general, $X_a$ is \emph{simpler} than $X$ (\eg, $X_a$ can be
finite) and allows interpretations $\den{-}_a$ of the language for
computing over-approximations of $\den{-}$, \ie,
$\gamma(\den{p}_a)\leq\den{p}$ for every (program) $p$ in the
language.

The adjunction between $X_a$ and $X$ gives a systematic way of
defining a $\den{-}_a$ from $\den{-}$.  For instance, if the language
is given by the BNF $p::=c \mid f(p)$, then an interpretation
$\den{-}$ in $X$ is uniquely determined by $\den{c}:X$ and a monotonic
map $\den{f}:X\to X$, and the \emph{abstract} interpretation in $X_a$
(computing over-approximations) is determined by taking
$\den{c}_a\defeq\alpha(\den{c}):X_a$ and
$\den{f}_a\defeq\alpha\circ\den{f}\circ\gamma:X_a\to X_a$. In static
analysis, the choice of $X_a$ and $\den{-}_a$ is a matter of
trade-offs between the cost of computing $\den{p}_a$ and the
information provided by $\gamma(\den{p}_a)$.

\paragraph{Spaces.}
To model more complex systems, \eg, continuous or hybrid
\cite{GST2009}, one may have to replace sets with more complex spaces.
For instance, in
\cite{Moggi_Farjudian_Duracz_Taha:Reachability_Hybrid:2018}, hybrid
systems are modelled by triples $(\State,F,G)$, where $\State$ is a
Banach space, and $F$ and $G$ are binary relations on
$\State$---called flow and jump relation, respectively---which
constrain how the system may evolve continuously (in time) and
discontinuously (instantaneously).  As in transition systems, the
relations $F$ and $G$ allow to model known unknowns.
Despite the increased complexity of models, it is still possible and
useful to move to the category of complete lattices and monotonic maps
and use it for analyses and abstract interpretations of these more
complex systems.

\paragraph{Imprecision.}
In defining reachability for hybrid systems, we realized the need to
cope with \emph{imprecision} (see also~\cite{franzle1999analysis}).
Thus, in~\cite{Moggi_Farjudian_Duracz_Taha:Reachability_Hybrid:2018},
we introduced \emph{safe and robust} reachability analysis.
In~\cite{MoggiFT-ICTCS-2019}, we use metric spaces to formalize the
notions of imprecision and robust analysis.
In a metric space $\State$, a level of imprecision $\delta>0$ means
that one cannot distinguish two points $s$ and $s'$ when their
distance $d(s,s')$ is less than $\delta$.  If one considers subsets
instead of points, and allows $\delta$ to become arbitrarily small,
then one cannot distinguish two subsets that have the same closure,
namely, $\forall \delta>0.B(S,\delta)=B(\cl{S},\delta)$, where
$B(S,\delta)$ is the open subset $\{s' \mid \exists s:S.d(s,s')<\delta\}$
and the closure $\cl{S}$ is the smallest closed subset containing $S$.
Therefore, one can replace $\pL(\State)$ with the complete lattice
$\cL(\State)$ of closed subsets, which is related to the former by the
adjunction%
\begin{tikzcd}[column sep = large]
  \cL(\State) \arrow[r, hook, "\top", "\gamma"' , yshift = -1.1ex] &
  \pL(\State) \arrow[l, twoheadrightarrow, "\alpha"' , yshift = 1.1ex],
\end{tikzcd}with $\gamma$ an inclusion map and $\alpha$ a surjective map.
This replacement is convenient, since the cardinality of $\cL(\State)$
can be smaller than that of $\pL(\State)$, \eg, when $\State$ is the
real line $\Real$.

\paragraph{Robustness.}
A monotonic map $A:\pL(\State)\to \pL(\State')$, with $\State$ and
$\State'$ metric spaces, is \emph{robust} when \emph{small} input
changes cause small output changes, \ie,
$\forall
S:\pL(\State).\forall\epsilon>0.\exists\delta>0.B(A(S),\epsilon)\leq
A(B(S,\delta))$.

When $A$ is robust, there is no \emph{loss of information} in
restricting to closed subsets, namely, there exists a unique monotonic
map $A_c:\cL(\State)\to \cL(\State')$ such that
$A_c\circ\alpha=\alpha\circ A$.  Thus, the focus
of~\cite{Moggi_Farjudian_Duracz_Taha:Reachability_Hybrid:2018,MoggiFT-ICTCS-2019}
is on analyses between complete lattices of closed subsets.
\cite{MoggiFT-ICTCS-2019} identifies sufficient (and almost necessary)
conditions to ensure that every monotonic map
$A:\cL(\State)\to \cL(\State')$ has a \emph{best robust
  approximation}, \ie, the biggest robust map
$\Box_R(A): \cL(\State)\to \cL(\State')$ such that $\Box_R(A)\leq A$ in
the lattice of monotonic maps with the point-wise order.

\paragraph{Scott continuity.}
We refer to~\cite{Gierz-ContinuousLattices-2003} for the definitions
of Scott continuous map, way-below relation $\ll$, and continuous
lattice. Restricting to compact metric spaces is mathematically
appealing, since in this case a monotonic map
$A:\cL(\State)\to \cL(\State')$ is robust exactly when it is Scott
continuous, and the complete lattices $\cL(\State)$ and $\cL(\State')$
are
$\omega$-continuous~\cite{Moggi_Farjudian_Duracz_Taha:Reachability_Hybrid:2018}.

A complete lattice $X$ is \emph{$\omega$-continuous} when it has a
countable \emph{base} $B$, \ie, a countable subset of $X$ such for
every $x:X$, the subset $B_x\defeq\{b:B \mid b\ll x\}$ is directed and
$x = \sup B_x$. Moreover, by fixing an enumeration $e$ of the base,
one can define when an element $x:X$ is \emph{computable}, namely,
when the set $\{n \mid e(n)\in B_x\}$ is a recursively enumerable
subset of $\Nat$.
The notion of computable can be extended to Scott continuous maps
between $\omega$-continuous lattices because these maps, under the
point-wise order, form an $\omega$-continuous lattice.

Ideally, we would like to focus on computable analyses, but we settle
for the broader class of Scott continuous analyses, since they are
better behaved.
For instance, every monotonic map $A:X\to X'$ between complete
lattices has a \emph{best Scott continuous approximation}
$\Box_S(A):X\to X'$, while there is no \emph{best computable
  approximation} of a monotonic map between (\emph{effectively given})
$\omega$-continuous lattices.

\paragraph{Related results.}
In~\cite{Moggi_Farjudian_Duracz_Taha:Reachability_Hybrid:2018}, we
define \emph{robustness} as a property of \emph{analyses}, \ie,
monotonic maps $A:\cL(\State_1)\to \cL(\State_2)$, where
$\State_i$ are metric spaces, and $\cL(\State_i)$ are the complete
lattices of closed subsets of $\State_i$, ordered by reverse
inclusion.
In the same paper, it was proven that:

\begin{itemize}
\item Robustness of $A$ amounts to continuity with respect to suitable
  $T_0$-topologies $\tau_R(\State_i)$ on the carrier sets of
  $\cL(\State_i)$, called \emph{robust topologies} (see
  Definition~\ref{def-top-S}). In general, these topologies depend on
  the metric structures of $\State_i$.

\item When $\State_i$ is compact, the topology $\tau_R(\State_i)$
  coincides with the \emph{Scott topology} $\tau_S(\State_i)$ on
  $\cL(\State_i)$. Thus, in this case, $\tau_R(\State_i)$ depends only
  on the \emph{topology} induced by the metric structure on
  $\State_i$.
\end{itemize}
\noindent
In particular, when both $\State_1$ and $\State_2$ are compact,
robustness and Scott continuity are equivalent properties of $A$, and
the complete lattices $\cL(\State_i)$ are $\omega$-continuous.
In~\cite{MoggiFT-ICTCS-2019}, we prove that every analysis
$A:\cL(\State_1)\to \cL(\State_2)$ has a \emph{best robust
  approximation} $\Box_R(A)$, when $\State_2$ is compact, with
$\Box_R(A)(C)$ given by $\bigcap\{A(C_\delta) \mid \delta>0\}$, where
the closed subset $C_\delta\defeq\cl{B(C,\delta)}$ is called
\emph{$\delta$-fattening} of~$C$.
When $\State_1$ is not compact, however, $\Box_R(A)$ may fail to be Scott
continuous, and $\cL(\State_1)$ may fail to be $\omega$-continuous.

\paragraph{Motivating examples.}
Examples of metric spaces that are not compact are Banach Spaces. In
applications, one usually considers closed bounded subsets of Banach
spaces. In finite-dimensional Banach spaces, all closed bounded
subsets are compact, but this fails in the infinite-dimensional case.
To motivate the need to go beyond compact subsets, we present some
examples of closed bounded subsets of infinite-dimensional Banach
spaces that are not compact:
\begin{itemize}
\item Probability distributions for a system with a countable set of
  states form a closed bounded subset of $\ell_1$, \ie, the
  Banach space of sequences $(x_n \mid n:\omega)$ in $\Real^\omega$ such that
  $\sum_{n:\omega}\absn{x_n}$ is bounded.

  More generally, probability distributions on a measurable space
  $(X,\Sigma)$ form a closed bounded subset of $ca(\Sigma)$, \ie,
  the Banach space of countably additive bounded signed measures on
  $\Sigma$.  This subset is not compact when the cardinality of
  $\Sigma$ is infinite.

\item Continuous maps from a compact Hausdorff space $X$ to a compact
  interval $[a,b]$ in $\Real$ form a closed bounded subset of $C(X)$,
  \ie, the Banach space of continuous maps from $X$ to $\Real$.  For
  instance, these maps could represent the height as a function of
  the position.

\item Closed bounded subsets of feature spaces arising from kernel
  methods in machine
  learning~\cite{Hofmann_et_al:Kernel_methods:2008}. Usually feature
  spaces are Hilbert spaces, whose carrier sets consist of real-valued
  maps. All Hilbert spaces with a countable orthonormal base are
  isomorphic to $\ell_2$, \ie, the Hilbert space of sequences
  $(x_n \mid n:\omega)$ in $\Real^\omega$ such that
  $\sum_{n:\omega}\absn{x_n^2}$ is bounded.

\item Closed bounded subsets of Sobolev spaces $W^{m,p}(\Omega)$, in
  which $\Omega \subseteq \Real^n$ is an open set. These sets commonly
  appear in solution of partial differential
  equations~\cite{Brezis:Functional_Analysis-Book:2011}.

\end{itemize}

\paragraph{Contribution.}
For simplicity, we consider analyses of the form
$A:\cL(\State)\to \cL(1)$, with $1$ denoting the one-point metric
space, although the results hold also when $1$ is replaces by a
compact metric space, \eg, a compact interval $[a,b]$ of the real
line. $\cL(1)$ is (isomorphic to) the two-point lattice $\Sigma$
and the complete lattice $\cL(\State)\to \Sigma$ is
isomorphic to that of upward closed subsets of $\cL(\State)$, ordered
by inclusion.

This paper proposes a way to reconcile robustness and Scott continuity
when $\State$ is not compact.
The general idea is to \emph{construct} an $\omega$-continuous lattice
$D$ related to $\cL(\State)$ by an adjunction
\begin{equation*}
  \begin{tikzcd}[column sep = large]
    \cL(\State) \arrow[r, "\top"', "\iota_*" , yshift =
    1.1ex] & D  \arrow[l, dashed, "\iota^*" , yshift = -1.1ex]
  \end{tikzcd}
\end{equation*}
such that the composite map $A'\circ \iota_*:\cL(\State)\to \Sigma$ is
robust whenever the map $A':D\to \Sigma$ is Scott
continuous. Therefore, given an analysis $A:\cL(\State)\to \Sigma$, we
can take the \emph{best Scott continuous approximation} $A'$ of
$A\circ \iota^*:D\to \Sigma$---in fact, any Scott continuous
approximation will do---and the composite map $A'\circ \iota_*$ is
guaranteed to be a robust approximation of $A$.

The $\omega$-continuous lattice $D$ that we construct is of the form
$\cL(\ol{\State})$, where $\ol{\State}$ is a compact Hausdorff space
given by the limit of an $\omega^{op}$-chain of compact metric spaces
related to $\State$ (see Theorem~\ref{thm:KMS}), and the adjunction
$\iota_*\vdash \iota^*$ is determined by a continuous map
$\iota:\State\to \ol{\State}$. Thus, by moving from $\State$ to
$\ol{\State}$, we gain compactness by giving up the metric structure.

In general, $\ol{\State}$ is not uniquely determined by $\State$,
although Theorem~\ref{thm:idem} provides some criteria to choose
the $\omega^{op}$-chain of compact metric spaces which determines
$\ol{\State}$.

\subsection*{Summary}
The rest of the paper is organized as follows:
\begin{itemize}
\item Section~\ref{sec:preliminaries} contains the mathematical
  preliminaries, where we fix notation and definitions. We will also
  present some basic results, usually without proofs, unless the
  results are not available in textbooks, in which case we provide
  proofs or pointers to other papers which include the relevant
  proofs. Most definitions are standard or taken from other
  papers. The only exception is the category $\Top_A$ of topological
  analyses (Definition~\ref{def-analyses}).

\item The main theoretical results are in
  Section~\ref{sec:main}. These include properties of idempotents and
  their splittings in a generic category $\C$ (\eg,
  Theorem~\ref{thm:idempotent}) and the construction
  (Theorem~\ref{thm:KMS}) of a continuous map
  $\iota:\State\to\ol{\State}$ relating a metric space $\State$ to a
  compact Hausdorff space $\ol{\State}$.

\item In Section~\ref{sec:examples}, we apply the results of
  Section~\ref{sec:main} to several examples of $\State$, namely:
  finite-dimensional Banach spaces $\ssp{m,p}$, infinite
  dimensional Banach spaces $\ssp{p}$ (\ie, sequence spaces), and
  closed unit balls $\cub{p}$ in sequence spaces.

\item In Section~\ref{sec:precision}, we investigate loss of precision
  when moving from the complete lattice $\cL(\State)$ to the
  $\omega$-continuous lattice $\cL(\ol{\State})$, when $\State$ is a
  closed unit ball $\cub{p}$.
  In Theorem~\ref{thm:preservation_of_precision} we characterize the
  closed subsets of $\ssp{p}$ (with $1 < p < \infty$) for which there
  is no loss of precisions as those that can be expressed as a
  non-empty intersections of finite unions of closed balls.

\item We conclude the paper with some remarks and suggestions for
  future work in Section~\ref{sec:concluding_remarks}.

\end{itemize}

\section{Mathematical Preliminaries}\label{sec:preliminaries}

In this section, we present the basic technical background---including
the notation---that will be used throughout the paper. We assume
standard terminology for topological and metric spaces. At times, we
may refer to a structure by its carrier set. For instance, for a
metric space $(\State, d)$, we may simply write `the metric space
$\State$'.

We use both $x \in X$ and $x:X$ to denote membership. A natural number
is identified with the set of its predecessors, {\ie}, $0 = \emptyset$
and $n = \set{0, \ldots, n-1}$, for any $n \geq 1$.
We write $\Nat$ or $\omega$ for the set of natural numbers.  When the
order matters $\omega$ denotes the set of natural numbers ordered by
inclusion, while $\Nat$ denotes the set of natural numbers with the
discrete order.
We write $(x_n \mid n:\omega)$ to denote a countable sequence, and
when the indexing set is clear from the context, we just write $(x_n
\mid n)$.

The powerset of a set $X$ is denoted by $\PS(X)$, $\subseteq$
denotes subset inclusion, and $\subset$ denotes strict subset
inclusion, {\ie}, $A \subset B \iff A \subseteq B \wedge A \neq
B$. Similarly, the finite powerset ({\ie}, the set of
finite subsets) of $X$ is denoted by $\PS_f(X)$, and $A\subseteq_f B$ denotes
that $A$ is a finite subset of $B$.

\subsection{Categories of Spaces}

\begin{table}[t]
\begin{equation*}
  \begin{tikzcd}[row sep = large, column sep = large]
    & \BS \arrow[r, hookrightarrow, "\bot", yshift = -1.1ex] \arrow[ d]
    & \NVS
    \arrow[l, dashed, yshift = 1.1ex]
    \arrow[ d]
    &
    &
    \text{normed vector space}
    \\
    \KMS \arrow[r, hookrightarrow]  \arrow[d]& \CMS
    \arrow[r, hookrightarrow, "\bot", yshift = -1.1ex]
    & \MS \arrow[l, dashed, yshift = 1.1ex]
    \arrow[d]
    &
    &
    \text{(extended) metric space}
    \\
    \KH \arrow[rr, hookrightarrow, "\bot", yshift = -1.1ex] &
    &
    \Haus
    \arrow[ll, dashed, yshift = 1.1ex]
    \arrow[r, hookrightarrow, "\bot", yshift = -1.1ex]
    &
    \Top_0 \arrow[l, dashed, yshift = 1.1ex]
    \arrow[d, "\vdash", xshift = -1.1ex]
    &
    \text{topological space}
    \\
    & & & \Set \arrow[u, dashed, xshift = 1.1ex]
  \end{tikzcd}
\end{equation*}
The notation%
\begin{tikzcd}
  {} \arrow[r] &  {}
\end{tikzcd}denotes a faithful \emph{forgetful} functor,%
\begin{tikzcd}
  {} \arrow[r, hookrightarrow] &  {}
\end{tikzcd}denotes the inclusion functor from a full sub-category, and $\vdash$
indicates the existence of a left adjoint to a functor.
The left adjoints from top to bottom and left to right are:
the Cauchy completion $\cl{X}$ (for $X$ in $\NVS$ or $\MS$),
the Stone-Cech compactification $\beta X$ (for $X$ in $\Haus$),
the Hausdorff reflection $HX$ (for $X$ in $\Top_0$),
the Discrete topology $DX$ (for $X$ in $\Set$).

\caption{Categories of Spaces}
\label{tab:spaces}
\end{table}

The spaces of interest for this paper are (extended) metric
spaces\footnote{We use extended metric spaces because they have better
  category-theoretic properties.} and Hausdorff spaces.  However, in
examples we restrict to Banach spaces, and some constructions extend
to arbitrary topological spaces.
Table~\ref{tab:spaces} summarizes the relations among
the following categories of spaces.

\begin{definition}[Categories of Spaces]\label{def-cat-spaces}\
\begin{itemize}
\item
  $\Top_0$ is the category of $T_0$-topological spaces $(X,\tau)$ and
  continuous maps.
  $\Haus$ and $\KH$ are the full sub-categories consisting of
  Hausdorff spaces (aka $T_2$-spaces) and compact Hausdorff spaces,
  respectively.
\item $\MS$ is the category of \textbf{extended metric spaces}
  $(X,d)$, \ie, the distance $d$ can be $\infty$, and \textbf{short
    maps}, \ie, maps $f: X_1 \to X_2$ such that
  $d_2(f(x),f(x'))\leq d_1(x,x')$ for $x,x':X_1$.
  There are other maps one can consider between (extended) metric
  spaces, in particular \textbf{isometries}, \ie, distance preserving
  maps.
  The forgetful functor%
  \begin{tikzcd}
  U:\MS \arrow[r] & \Haus
  \end{tikzcd}maps a distance $d$ on $X$
  to the $T_2$-topology $\tau_d$ on $X$ generated by the open balls.
  $\CMS$ and $\KMS$ are the full sub-categories of $\MS$ consisting of
  Cauchy complete extended metric spaces and compact extended metric spaces,
  respectively.  The objects in $\KMS$ are exactly the extended metric spaces
  whose underlying topological spaces are compact.

\item
  $\NVS$ is the category of normed vector spaces
  $(X,\cdot,+,\norm{-})$ and short linear maps.
  The forgetful functor $U:\NVS \to \MS$ maps a normed vector space to
  the metric space with (the same carrier and) distance
  $d(x',x)\defeq\norm{x'-x}$.
  $\BS$ is the full sub-category of $\NVS$ consisting of Banach
  spaces. The objects in $\BS$ are exactly the normed vector spaces
  whose underlying metric spaces are complete.
\end{itemize}
\end{definition}

The following theorems recall some properties of categories and functors in
Table~\ref{tab:spaces}.
\begin{theorem}\label{thm-cat-KMS}
  The categories in the following diagram have finite limits and
  finite sums, and the functors preserve them:
  \begin{equation*}
    \begin{tikzcd}
      \KMS \arrow[r, hook] & \CMS \arrow[ r, hook] & \MS \arrow[r] &\Haus
    \end{tikzcd}
  \end{equation*}
  The categories $\CMS$, $\MS$ and $\Haus$ have also small limits and
  small colimits.
\end{theorem}
For the existence of sums and infinitary products it is essential to
use extended metric spaces.

\begin{theorem}\label{thm-cat-KH}
  The categories in the following diagram have small limits and finite
  sums, and the functors preserve them:
  \begin{equation*}
    \begin{tikzcd}
      \KH \arrow[r, hook] & \Haus \arrow[r, hook]&\Top_0
    \end{tikzcd}
  \end{equation*}
  The categories have also small colimits.
\end{theorem}

\subsection{Categories of Analyses}

We define an analysis as a monotonic map between complete lattices.
However, we need to consider further properties of analyses, that
(with the exception of computability) can be defined as continuity
with respect to suitable topologies on (the carrier set of) complete
lattices.  For this reason, we introduce the category $\Top_A$ of
\emph{topological analyses}, which \emph{refines} the category $\Po_A$
of \emph{analyses} (see Table~\ref{tab:Top_Po_Categories}).

\begin{table}[t]
  \begin{equation*}
  \begin{tikzcd}[row sep = large, column sep = large]
    \Top_A \arrow[r, hookrightarrow] \arrow[d, xshift = 1.1ex]& \Top_0
    \arrow[d, "U", xshift = 1.1ex]
    & \text{topologies}\\
    \Po_A \arrow[r, hookrightarrow] \arrow[u, hook', swap, "\dashv",
    xshift = -1.1ex]& \Po \arrow[u, hook', swap, "\dashv",
    xshift = -1.1ex, "A"'] & \text{posets}
  \end{tikzcd}
\end{equation*}
The notation
\begin{tikzcd}
  {} \arrow[r] & {}
\end{tikzcd}
denotes a faithful \emph{forgetful} functor,
\begin{tikzcd}
  {} \arrow[r, hook] & {}
\end{tikzcd}
denotes the inclusion functor from a full sub-category, and $\vdash$
indicates the existence of a left adjoint to a functor.

\caption{Poset-enriched categories}
\label{tab:Top_Po_Categories}
\end{table}

\begin{definition}[Category of Analyses]\label{def-analyses}\
  \begin{itemize}
  \item
    $\Po$ is the category of posets and monotonic maps.
  \item
    The forgetful functor%
    \begin{tikzcd}
      U: \Top_0 \arrow[r] & \Po
    \end{tikzcd}maps a $T_0$-topology $\tau$ on $X$ to the
    \textbf{specialization order} $\leq_\tau$ on $X$, \ie,
    $x\leq_\tau y\defiff\forall O:\tau.x\in O\implies y\in O$.

  \item
    The inclusion functor%
    \begin{tikzcd}
      A:\Po \arrow[r, hook] &\Top_0
    \end{tikzcd}maps a poset $\leq$ on $X$ to the \textbf{Alexandrov topology}
    $\tau_\leq$ on $X$ consisting of the upward closed subsets, \ie,
    $O\in\tau_\leq\defiff\forall x,y:X.x\in O\land x\leq y\implies
    y\in O$.

    %
  \item
    $\Po_A$, the category of \textbf{analyses}, is the full
    sub-category of $\Po$ consisting of complete lattices.
  \item
    $\Top_A$, the category of \textbf{topological analyses}, is the
    full sub-category of $\Top_0$ consisting of $T_0$-spaces whose
    specialization order is a complete lattice.
\end{itemize}
\end{definition}
\begin{theorem}\label{thm-po-top}\
  \begin{enumerate}
  \item
    The categories $\Po$ and $\Top_0$ are $\Po$-enriched and have small
    limits and small colimits.
  \item
    The functors $U$ and $A$ are $\Po$-enriched, and $A$ is left adjoint
    to $U$.
  \item
    The functor $U$ preserves small limits and small sums.

  \item
    The functor $A$ preserves finite limits and small
    colimits.
  \end{enumerate}
\end{theorem}

\begin{proof}
  The $\Po$-enrichment of $\Po$ is given by its cartesian closed
  structure. Since $U$ is faithful, $\Top_0(X,Y)$ is a subset of
  $\Po(UX,UY)$ and inherits the $\Po$-enrichment.
  It is easy to prove that $\Top_0(AX,Y)=\Po(X,UY)$. Thus, $A$ is left
  adjoint to $U$ also as $\Po$-enriched functors.
\end{proof}

\begin{corollary}\label{cor-po-top-A}\
  \begin{enumerate}
  \item
    The category $\Po_A$ is $\Po_A$-enriched and has small products.
  \item The category $\Top_A$ is $\Po$-enriched and has small
    products.

  \item
    The $\Po$-enriched functors $U$ and $A$ restrict to functors
    between $\Top_A$ and $\Po_A$.
  \end{enumerate}
\end{corollary}
\begin{proof}
  The $\Po_A$-enrichment of $\Po_A$ is given by its cartesian closed
  structure.  The other claims are easy consequences of
  Theorem~\ref{thm-po-top} and the definition of $\Top_A$.
\end{proof}

\begin{theorem}[Topologies on a poset]
  \label{thm:spec_order_comp_lat}
  Given a partial order $\leq$ on $X$, the set of $T_0$-topologies on
  $X$ with specialization order $\leq$ ordered by
  reverse inclusion forms a complete lattice
  $\Top(\leq)$, where:
  \begin{itemize}
  \item
    the least element $\tau_\bot$ is the Alexandrov topology $\tau_\leq$.
  \item
    the top element $\tau_\top$ is the topology generated by
    the set $\setbarNormal{\not\downarrow y}{y:X}$, where $\not\downarrow y\defeq\{x:X \mid x\not\leq y\}$.
  \item
    the nonempty sups are given by intersection.
  \end{itemize}
  Moreover, when $X$ is finite $\Top(\leq)$ is trivial, \ie,
  $\tau_\bot=\tau_\top$.
\end{theorem}

\begin{proof}
  It is easy to show that $(X,\tau_\top)$ and $(X,\tau_\bot)$ are
  $T_0$-spaces with specialization order $\leq$.
  Given a $T_0$-topology $\tau$ on $X$ with specialization order
  $\leq$ we have $\tau_\top\subseteq\tau\subseteq\tau_\bot$, because:
  \begin{itemize}
  \item
    each $O:\tau$ is upward closed, by definition of $\leq_\tau$,
  \item
    each $\not\downarrow y$ is in $\tau$, because $\not\downarrow
    y=\bigcup\{O:\tau \mid y\not\in O\}$.
  \end{itemize}
  Therefore, $\tau_\top$ and $\tau_\bot$ are respectively the top and
  bottom element in $\Top(\leq)$.
  Since the topologies on a set $X$ (ordered by reverse inclusion)
  form a complete lattice, with (nonempty) sups given by
  intersections, so do the topologies $\tau$ on $X$ such that
  $\tau_\top\subseteq\tau\subseteq\tau_\bot$.  Moreover, such
  topologies are $T_0$, because $\tau_\top$ is.

  For every $x:X$ we have $\uparrow x=\bigcap\{\not\downarrow
  y \mid y:X\land x\not\leq y\}$.  When $X$ is finite, the rhs of the
  equality is a finite intersection of open sets in $\tau_\top$, thus
  $\uparrow x\in\tau_\top$, and therefore $\tau_\bot\subseteq\tau_\top$.
\end{proof}

When $\leq$ is a complete lattice (\ie, an object in $\Po_A$), the
topologies in $\Top(\leq)$ are objects in $\Top_A$.

\subsection{Adjunctions and Best Approximations}

A key property of categories of analyses is poset-enrichment,
which provides a qualitative criterion for comparing analyses between
two complete lattices, and allows to define \emph{adjunctions}
(aka Galois connections) between two complete lattices.

\begin{definition}[Adjunction]\label{def-adj}\
  An \textbf{adjunction} in a $\Po$-enriched category $\C$, notation
  $f\dashv g$, is a pair of maps%
  \begin{tikzcd}
    X \arrow[r, "f" description, yshift = -1.1ex] & Y \arrow[l ,
    "g" description, yshift = 1.1ex]
  \end{tikzcd}in $\C$ such that $f\circ g\leq\id_Y$ and
  $g\circ f\geq\id_X$.  The maps $f$ and $g$ are called left- and
  right adjoint, respectively. Moreover, any one of these two maps
  uniquely determines the other.
\end{definition}

\begin{remark}\label{rmk-adj-po}
A characterization of adjunctions in $\Po$ is $f\dashv g$ iff $\forall
x:X,y:Y.x\leq_X g(y)\iff f(x)\leq_Y y$.  This characterization implies
that in $\Po$ left adjoints preserve sups and (dually) right adjoints
preserve infs.
\end{remark}

\begin{theorem}\label{thm-CPL_m}
  The $\Po$-enriched categories $\Po_A$ and $\Top_A$ have limits of
  $\omega^{op}$-chains of right adjoints.
\end{theorem}


\begin{proof}
  First, we prove the property for $\Po_A$. Given an
  $\omega^{op}$-chain $(p_n:D_{n+1}\to D_n \mid n)$ of right adjoints
  in $\Po_A$, its limit in $\Po$ is the subset of
  $\prod_{n}\absn{D_n}$ given by
  $\absn{D}\defeq\{d \mid \forall n.d_n=p_n(d_{n+1})\}$ with the
  point-wise order $\leq_D$.
  Since right adjoints preserve infs, infs in $D$ exist and are
  computed point-wise. Thus, $D:\Po_A$ and the maps $\pi_n:D \to D_n$
  with $\pi_n(d)=d_n$ form a limit cone (and preserve infs).

  Given an $\omega^{op}$-chain $(p_n:D_{n+1} \to D_n \mid n)$ of right
  adjoints in $\Top_A$, its limit in $\Top_0$ is the sub-space of
  $\prod_{n}D_n$ corresponding to the subset
  $\absn{D}\defeq\{d \mid \forall n.d_n=p_n(d_{n+1})\}$, and the maps
  $\pi_n:D \to D_n$ with $\pi_n(d)=d_n$ form a limit cone. Since
  $U:\Top_0 \to \Po$ is $\Po$-enriched and preserves limits, we have
  that $(Up_n \mid n)$ is an $\omega^{op}$-chain of right adjoints in
  $\Po_A$ and $(U\pi_n \mid n)$ is a limit cone in $\Po$.  By the
  result for $\Po_A$, we have $UD:\Po_A$. Hence, $D:\Top_A$.
\end{proof}
A similar result holds if right adjoints are replaced by left adjoints
(\ie, $\Po_A$ and $\Top_A$ have limits of $\omega^{op}$-chains of left
adjoints).
We recall further properties of adjunctions in
$\Po$-enriched categories (and in $\Po$).  Each of these properties
has a dual, that we do not state explicitly.

\begin{proposition}\label{thm-adj-mono}
  If $f\dashv g$ in a $\Po$-enriched category $\C$ and $f:X\to Y$ is
  mono, then $g\circ f=\id_X$.
\end{proposition}

\begin{proof}
  Since $f\dashv g$ one has $f\circ g\circ f=f$. When $f$ is mono,
  this implies $g\circ f=\id_X$.
\end{proof}

In other words, if a left adjoint is mono, then it is
split mono and its right adjoint is split epi.

\begin{proposition}\label{thm-adj-sups}
  If $X$ is a complete lattice, then $f:X \to Y$ is a left adjoint in
  $\Po$ iff $f$ preserves sups.
\end{proposition}

\begin{proof}
By Remark~\ref{rmk-adj-po}, left adjoints in $\Po$ preserve sups, so it remains
to prove the right-to-left implication.  Since $X$ is a complete lattice,
$g(y)\defeq\sup\{x \mid f(x)\leq_Y y\}$ is well-defined.  We prove that
$f\dashv g$, or equivalently $x\leq_X g(y)\iff f(x)\leq_Y y$.
The left-to-right implication follows from:
\begin{itemize}
\item $x\leq_X g(y)\implies$ because $f$ preserves sups
\item $f(x)\leq_Y f(g(y))=\sup\{f(x) \mid f(x)\leq_Y y\}\leq_Y y$.
\end{itemize}
The right-to-left implication follows from:
\begin{itemize}
\item $f(x)\leq_Y y\implies$ by definition of $g$
\item $x\leq_X\sup\{z \mid f(z)\leq_Y y\}=g(y)$.
\end{itemize}
\end{proof}

\begin{proposition}\label{thm-adj-cpl}
  If $Y$ is a complete lattice and $X$ is a sub-poset of $Y$, then the
  inclusion $f:X \to Y$ is a left adjoint in $\Po$ iff $X$ is a
  complete lattice and sups in $X$ are computed as in $Y$ (\ie, $f$
  preserves sups).
\end{proposition}
\begin{proof}
  The right-to-left implication follows from Proposition~\ref{thm-adj-sups}.
  For the other implication, consider the right adjoint $g$ to $f$.
  By Remark~\ref{rmk-adj-po}, $f$ preserves sups and $g$ preserves
  infs.  Moreover, $X$ has all infs (\ie, is a complete lattice),
  because $g\circ f=\id_X$ (by Proposition~\ref{thm-adj-mono}) and
  $\inf D=g(\inf f(D))$ for any subset $D$ of $X$.
  \end{proof}

  \begin{definition}[Best Approximation]
    \label{def-ba}
    Given a subset $X$ of a poset $Y$, $x:X$ is the \textbf{best
      $X$-approximation} of $y:Y$ iff
    $\forall x':X.x'\leq x\iff x'\leq y$.
  \end{definition}

  A subset $X$ of a poset $Y$ can be identified with the sub-poset of
  $Y$ with carrier $X$ and the partial order inherited from $Y$. Then,
  the inclusion $f:X \to Y$ is a left adjoint in $\Po$ exactly when
  every $y:Y$ has a best $X$-approximation, and the right adjoint to
  $f$ maps $y$ to its best $X$-approximation.
  When $Y$ is a complete lattice, Proposition~\ref{thm-adj-cpl}
  characterizes the subsets $X$ of $Y$ for which every $y:Y$ has a
  best $X$-approximation.

  We are interested in best $X$-approximations of analyses in
  $\Po_A(\leq_1,\leq_2)$ for $X$ of the form $\Top_A(\tau_1,\tau_2)$,
  where $\leq_i$ is the specialization order of the topology $\tau_i$,
  \ie, $\leq_i=U(\tau_i)$. For existence of best $X$-approximations,
  the poset $\Top_A(\tau_1,\tau_2)$ must be a complete lattice with
  sups computed as in $\Po_A(\leq_1,\leq_2)$.

\begin{example}
  Given a complete lattice $\leq$ (on a set $X$), we can define two
  topologies in $\Top(\leq)$:
  \begin{enumerate}
  \item the \textbf{Scott topology} $\tau_S(\leq)$ consists of
    upward closed subsets $O$ of $X$ such that $\sup D\in O\implies \exists
    d:D.d\in O$ for any directed subset $D$ of $X$, equivalently, $\sup
    S\in O\implies\exists S_0\subseteq_f S.\sup S_0\in O$ for any subset
    $S$ of $X$.
  \item the \textbf{$\omega$-topology} $\tau_\omega(\leq)$ consists of
    upward closed subsets $O$ of $X$ such that
    $\sup D\in O\implies\exists d:D.d\in O$ for any $\omega$-chain $D$ in
    $X$, equivalently, $\sup S\in O\implies\exists S_0\subseteq_f S. \sup S_0\in O$ for any countable subset $S$ of $X$.
  \end{enumerate}
  Clearly, $\tau_S(\leq)\subseteq\tau_\omega(\leq)$.  Given a map
  $f:\Po_A(\leq_1,\leq_2)$, there is an order-theoretic
  characterization of continuity with respect to these two topologies,
  namely:
  \begin{itemize}
  \item $f$ is \textbf{Scott continuous}, \ie,
    $f:\Top_A(\tau_S(\leq_1),\tau_S(\leq_2))$, exactly when $f$
    preserves sups of directed sets.
  \item $f$ is \textbf{$\omega$-continuous}, \ie,
    $f:\Top_A(\tau_\omega(\leq_1),\tau_\omega(\leq_2))$, exactly when
    $f$ preserves sups of $\omega$-chains.
  \end{itemize}
  These characterizations imply that
  $\Top_A(\tau_S(\leq_1),\tau_S(\leq_2))\subseteq
  \Top_A(\tau_\omega(\leq_1),\tau_\omega(\leq_2))\subseteq
  \Po_A(\leq_1,\leq_2)$ and that these subsets are closed with respect
  to sups computed in $\Po_A(\leq_1,\leq_2)$. Therefore, every
  analysis $f:\Po_A(\leq_1,\leq_2)$ has a best Scott-continuous
  approximation $\Box_S(f)$, and a best $\omega$-continuous
  approximation $\Box_\omega(f)$, with
  $\Box_S(f)\leq\Box_\omega(f)\leq f$.
\end{example}

We introduce two sub-categories of $\Po_A$, related to the example
above, which can be viewed also as full sub-categories of $\Top_A$, by
mapping a complete lattice with order $\leq$ to the Scott topology
$\tau_S(\tau)$.
\begin{definition}[Category of Continuous Lattices~\cite{Compendium:Book:1980}]\label{def-CL}\
  \begin{itemize}
  \item
    $\CL$ is the category of continuous lattices, \ie, every element
    $x$ in the lattice is the sup of the directed set formed by the
    elements way-below $x$, and Scott continuous maps.
  \item
    $\wCL$ is the full sub-category of $\CL$ whose objects are
    $\omega$-continuous lattices, \ie, continuous lattices with a
    countable subset $B$ (called a \emph{base}) such that every
    element $x$ in the lattice is the sup of the directed set formed
    by the elements in the base way-below $x$.
\end{itemize}
\end{definition}

\begin{proposition}\label{thm-CL}\
The $\Po$-enriched categories in the following diagram have finite
products and limits of $\omega^{op}$-chains of right adjoints and the
functors preserve them:
\begin{equation*}
\begin{tikzcd}
  \wCL \arrow[r, hook] &\CL \arrow[r, hook] &\Top_A \arrow[r] &\Po_A.
\end{tikzcd}
\end{equation*}
Moreover, the categories $\wCL$ and $\CL$ have exponentials and the
functor%
\begin{tikzcd}
  \wCL \arrow[r, hook] &\CL
\end{tikzcd}preserves them, and every $\omega$-continuous map between
$\omega$-continuous lattices is necessarily Scott continuous.
\end{proposition}

\begin{proof}
  For finite products in $\wCL$ and $\CL$,
  see~\cite[Proposition~3.2.4]{AbramskyJung94-DT}. Exponentials in
  $\CL$ are discussed in~\cite[Section~II-4]{Compendium:Book:1980}. Scott
  continuity of $\omega$-continuous maps between $\omega$-continuous
  lattices is proven in~\cite[Proposition~2.2.14]{AbramskyJung94-DT}.
\end{proof}

\subsection{From Spaces to Complete Lattices}
\label{sec:form_spaces_to_lattices}

Given a topological space $\State$, the set of closed subsets of
$\State$, ordered by reverse inclusion, forms a complete lattice
$\cL(\State)$, with sups given by intersection. We introduce several
topologies on these complete lattices, but first we give the main
properties of $\cL$ as a functor from $\Haus$ to $\Po_A$.
\begin{definition}\label{def-cL}
  The functor%
  \begin{tikzcd}
    \cL:\Haus \arrow[r] & \Po_A
  \end{tikzcd}is defined as follows:
  \begin{itemize}
    \item $\cL(\State)$ is the complete lattice
      of closed subsets of $\State$ under reverse inclusion.
    \item If $f:\Haus(\State,\State')$, then the map $\cL(f)\defeq
      f_*:\Po_A(\cL(\State),\cL(\State'))$ is given by
      $f_*(C)=\cl{f(C)}$, \ie, it maps $C$ to the closure of the image
      of $C$ along $f$.
    \item[] There is also a contravariant version, whose action on
      maps $f^*:\Po_A(\cL(\State'),\cL(\State))$ is given by
      $f^*(C')=f^{-1}(C')$, \ie, it maps $C'$ to the inverse image of
      $C'$ along $f$.
  \end{itemize}
\end{definition}

\begin{theorem}\label{thm-cL1}
  If%
  \begin{tikzcd}
    \State \arrow[r, "f" description]&\State'
  \end{tikzcd}in $\Haus$, then $f^*\dashv f_*$ in $\Po_A$.
\end{theorem}

\begin{proof}
  See~\cite[Example~5.5]{Moggi_Farjudian_Duracz_Taha:Reachability_Hybrid:2018}.
\end{proof}

\begin{definition}[Topologies on $\cL(\State)$~\cite{Moggi_Farjudian_Duracz_Taha:Reachability_Hybrid:2018}]
  \label{def-top-S}\
  \begin{enumerate}
  \item   Given a Hausdorff space $\State$, we write $\OS(\State)$ for the set
  of open subsets of $\State$, $\CS(\State)$ for the set of closed
  subsets of $\State$, and $\uparrow S$ for the set of $C:\CS(\State)$
  such that $C\subseteq S$, where $S$ is a subset of $\State$.
  The \textbf{Upper topology} $\tau_U(\State)$ is the topology on
  $\CS(\State)$ such that $U:\tau_U(\State)\defiff\forall C:U.\exists
  O:\OS(\State).C:\uparrow O\subseteq U$.

  \item Given an extended metric space $\State$, we write
    $B(x,\delta):\OS(\State)$ for the open ball with center $x:\State$
    and radius $\delta>0$, and $B(S,\delta):\OS(\State)$ for the union
    of open balls $B(x,\delta)$ with $x:S$, where $S$ is a subset of
    $\State$.
  The \textbf{Robust topology} $\tau_R(\State)$ is the topology on
  $\CS(\State)$ such that $U:\tau_R(\State)\defiff\forall C:U.\exists
  \delta>0.C:\uparrow B(C,\delta)\subseteq U$.

  \item   For uniformity we write $\tau_S(\State)$ and $\tau_A(\State)$ for
  the Scott and Alexandrov topologies on $\cL(\State)$, and for
  conciseness we write $\mathbb{A}_{XY}(\State_1,\State_2)$ for the
  poset $\Top_A(\tau_X(\State_1),\tau_Y(\State_2))$.
  \end{enumerate}
\end{definition}
We call Upper topology what is better known as upper Vietoris topology.

\begin{theorem}\label{thm-top-S}\
  \begin{enumerate}
  \item If $\State:\Haus$, then $\tau_U(\State)$ is in $\Top(\leq)$, where $\leq$ is the partial
  order on $\cL(\State)$.
  \item \label{item:KH_CL} If $\State:\KH$, then $\cL(\State):\CL$ and
    $\tau_U(\State)=\tau_S(\State)$.
  \item If $\State:\MS$, then
    $\tau_S(\State)\subseteq\tau_R(\State)\subseteq\tau_U(\State)$.
    Therefore, $\tau_S(\State)=\tau_R(\State)=\tau_U(\State)$ when
    $\State:\KMS$.
  \item If $\State:\KMS$ is finite, then $\cL(\State)$ is finite and
    $\tau_S(\State)=\tau_A(\State)$.
  \end{enumerate}
\end{theorem}
\begin{proof}\
  \begin{enumerate}
  \item We prove that $\tau_\top\subseteq
    \tau_U(\State)\subseteq\tau_\bot=\tau_A(\State)$ (see
    Theorem~\ref{thm:spec_order_comp_lat}).
    $\tau_U(\State)\subseteq\tau_A(\State)$, because the open subsets
    of the upper topology are upward closed with respect to reverse
    inclusion.

    To prove $\tau_\top\subseteq\tau_U(\State)$ we show that
    $\not\downarrow C\defeq\{C':\CS(\State) \mid C\not\subseteq C'\}$
    is in $\tau_U(\State)$ when $C:\CS(\State)$.  If $C\not\subseteq
    C'$, then there exists $x$ in $C-C'$.  But every singleton is closed
    in $\State$ (because $\State:\Haus$), thus the complement $O$ of
    the singleton $\{x\}$ is open and $C'\in\uparrow
    O\subseteq\not\downarrow C$.

  \item Follows from~\cite[Proposition~3.3]{Edalat95:DT-fractals}.
  \item Follows
    from~\cite[Lemma~A.3]{Moggi_Farjudian_Duracz_Taha:Reachability_Hybrid:2018}.
  \item If $\State$ is finite, then $\cL(\State)$ is also
    finite. Hence, $\Top(\leq)$ contains only one topology.
  \end{enumerate}
\end{proof}

\begin{theorem}\label{thm-cL0}
  If $f:\Haus(\State,\State')$, then
  $f^*:\mathbb{A}_{SS}(\State',\State)$ and
  $f_*:\mathbb{A}_{UU}(\State,\State')$.
\end{theorem}
\begin{proof}
  Since the map $f^*$ is a left adjoint in $\Po_A$, it preserves all
  sups. Thus, it is Scott continuous.  The map $f_*$ is Upper
  continuous, because for any $C:\CS(\State)$ and $O':\OS(\State')$ we
  have
  $\cl{f(C)}\subseteq O'\implies f(C)\subseteq O'\implies C\subseteq
  f^{-1}(O')$ and $f^{-1}(O'):\OS(\State)$ by continuity of $f$.
\end{proof}

\begin{theorem}\label{thm-cL3}
  The functor%
  \begin{tikzcd}
    \cL:\Haus \arrow[r] &\Po_A
  \end{tikzcd}%
  restricted to $\KH$ factors through $\CL$, and when restricted to
  $\KMS$ factors through $\wCL$.
\end{theorem}

\begin{proof}
  From Theorem~\ref{thm-top-S} we know that for any $X:\KH$, the
  lattice $\cL(X)$ is continuous. The fact that for any $X:\KMS$, the
  lattice $\cL(X)$ is $\omega$-continuous is a straightforward
  consequence. It remains to show that, if $f:\KH(X,Y)$, then $f_*$ is
  Scott continuous. But this also follows from item~\ref{item:KH_CL}
  of Theorem~\ref{thm-top-S} and Theorem~\ref{thm-cL0}.
\end{proof}

\begin{theorem}\label{thm-cL2}
  The functor%
  \begin{tikzcd}
    \cL:\KH \arrow[r] & \Po_A
  \end{tikzcd}%
  preserves limits of $\omega^{op}$-chains.
\end{theorem}

\begin{proof}
  Given an $\omega^{op}$-chain $(p_n:X_{n+1} \to X_n \mid n)$ in
  $\KH$, its limit in $\KH$ (and $\Haus$) is the sub-space $X$ of
  $\prod_n X_n$ such that
  $\absn{X}=\{x \mid \forall n. x_n=p_n(x_{n+1})\}$.
  By Theorem~\ref{thm-CPL_m}, the limit of $(\cL(p_n) \mid n)$ in
  $\Po_A$ (and $\CL$) is the sub-poset $D$ of $\prod_nD_n$ such that
  $\absn{D}=\{d \mid \forall n.d_n=\cl(p_n)(d_{n+1})\}$, where
  $D_n\defeq\cL(X_n)$.  But for spaces in $\KH$ compact subsets and
  closed subsets coincide, thus $\cL(p_n)(C)=p_n(C)$, \ie, the image
  of $C$ along $p_n$.

  By the universal property of $D$ there exists a unique map
  $\phi:\cL(X) \to D$ in $\Po_A$ such that:
  \begin{equation*}
  \begin{tikzcd}[column sep = large]
    \cL(\pi_n:X \arrow[r] & X_n) = \cL(X)\arrow[r, "\phi" description]
    & D \arrow[r, "\pi_n" description] & D_n,
  \end{tikzcd}
\end{equation*}
namely, $\phi(C)=(\pi_n(C) \mid n)$ for every $C:\cL(X)$.  Moreover,
$\phi$ preserves infs, since the $\cL(\pi_n)$ are right adjoints and
preserve infs. Therefore, $\phi$ has a left adjoint
$\phi':D \to \cL(X)$, namely $\phi'(d)=\bigcap_n\pi_n^*(d_n)$.
  We prove that $\phi'$ is the inverse of $\phi$.
  Because of the adjunction $\phi'\dashv\phi$ we have:
  \begin{enumerate}
  \item $\forall n.\forall
    d:D.\phi(\phi'(d))_n=\pi_n(\bigcap_n\pi_n^*(d_n))\subseteq d_n$,
    and
  \item $\forall
    C:\cL(X).C\subseteq\phi'(\phi(C))=\bigcap_n\pi_n^*(\pi_n(C))$.
  \end{enumerate}

  Given $d:D$, we have (by the Axiom of Choice) $\forall n.\forall
  x:d_n.\exists y:d_{n+1}.x=p_n(y)$, which implies $\forall n.\forall
  x:d_n.\exists y:X.x=y_n\land(\forall i.y_i\in d_i)$.  But
  $\{y:X \mid \forall n.y_n\in d_n\}$ is another way to denote
  $\bigcap_n\pi_n^*(d_n):\cL(X)$, thus the first inclusion is an
  equality.

  For the second inclusion, consider an $x\in\phi'(\phi(C))$, \ie,
  $\forall n.x_n\in\pi_n(C)$, we prove that $x\in C$.
  Since $C$ is closed and a base for the topology on $X$ is given by
  the subsets $[O]_n=\{y:X \mid y_n\in O\}$ with $O:\OS(X_n)$, it
  suffices to prove that $\forall n.\forall
  O:\OS(X_n).x\in[O]_n\implies\exists y:C.y\in[O]_n$.
  But $x_n\in\pi_n(C)$ means that $x_n=y_n$ for some $y\in C$, thus
  $x\in [O]_n\iff x_n\in O\iff y_n\in O\iff y\in [O]_n$.

  \end{proof}


\section{Main Results}
\label{sec:main}

Ideally, given a metric space $\State$ (or more generally, an extended
metric space), we would like to find a \emph{compactification}
$\ol{\State}$ of $\State$ such that the complete lattice
$\cL(\ol{\State})$ is $\omega$-continuous.\footnote{For this, it
  suffices that the topology on $\ol{\State}$ has a countable base.}
We establish a weaker result, namely, given an $\omega$-chain
$(g_n \mid n)$ of short idempotents (with certain additional
properties) on a metric space $\State$, we define a compact Hausdorff
space $\ol{\State}$ with a countable base and a continuous map
$\iota:\State \to \ol{\State}$ such that the monotonic map
$\cL(\iota):\cL(\State) \to \cL(\ol{\State})$ is continuous when
$\cL(\State)$ is equipped with the Robust topology and
$\cL(\ol{\State})$ is equipped with the Scott topology.
In general, $\ol{\State}$ is not a compactification of $\State$, nor
is it \emph{uniquely} determined (up to isomorphism) by $\State$, as
it depends on the choice of $(g_n \mid n)$.

The result above follows from Theorem~\ref{thm:KMS}, which is
applicable under more general assumptions than having an
$\omega$-chain $(g_n \mid n)$ of short idempotents.
Another result, Theorem~\ref{thm:idem}, gives sufficient conditions to
ensure that $\iota:\State\to\ol{\State}$ is both mono and epi, which
is as close as we can get to have that $\ol{\State}$ is a
compactification of $\State$.

\subsection{Idempotents, Embeddings and Projections}
\label{sec:idempotents}

In this section, we establish general properties of idempotents, split
monos and split epis, that in this paper we call embeddings and
projections, respectively.

\begin{definition}[idempotents \& co]
  In a category $\C$:
  \begin{enumerate}
  \item $g$ is an \textbf{idempotent} on $X$ $\defiff$%
    \begin{tikzcd}
      X \arrow[r, "g" description]& X
    \end{tikzcd}
    and $g\circ g=g$.
  \item given two idempotents $g_1$ and $g_2$ on $X$ we write
    $g_1\leq g_2\defiff g_1\circ g_2 = g_1 = g_2\circ g_1$.
  \item $(e,p)$ is an \textbf{e-p pair} from $X$ to $Y$, notation
    \begin{tikzcd}
      (e,p):X \arrow[r, Rightarrow]& Y \defiff
      X \arrow[r, "e" description, yshift = -1.1ex] & Y \arrow[l, "p"
      description, yshift = 1.1ex]
    \end{tikzcd}
    and $p\circ e=\id_X$.
  \item $(e,p)$ is a \textbf{splitting} of $g$ $\defiff$ $(e,p)$ is an
    e-p pair and $g=e\circ p$.
  \item
    \textbf{idempotents split} (in $\C$) $\defiff$ every idempotent
    (in $\C$) has a splitting.
  \end{enumerate}
  Given an e-p pair $(e,p)$, we call $e$ \textbf{embedding} and $p$
  \textbf{projection}. In general, $e$ and $p$ do not determine each other.
  $\C_{ep}$ denotes the category whose arrows are e-p pairs and composition
  $(e_2,p_2)\circ(e_1,p_1)$ is given by $(e_2\circ e_1,p_1\circ p_2)$.
  The forgetful functors%
  \begin{tikzcd}
    E:\C_{ep}\arrow[r] &\C
  \end{tikzcd}
  and%
  \begin{tikzcd}
    P:\C_{ep}\arrow[r] & \C^{op}
  \end{tikzcd}
  map the e-p pair $(e,p)$ to the embedding $e$ and the projection
  $p$, respectively.
\end{definition}

The notions above are \emph{absolute}, \ie, they are preserved by
functors, because they are defined only in terms of composition and
identities.  For instance, if%
\begin{tikzcd}
  F:\C\arrow[r] &\C'
\end{tikzcd}
is a functor and $g$ is an idempotent on $X$ in $\C$, then $Fg$ is an
idempotent on $FX$ in $\C'$.
As is customary in Category Theory, a definition or result given for a
generic category $\C$ can be recast in the dual category $\C^{op}$.

\begin{proposition}[Duality]\label{thm:idem-dual}\
  \begin{enumerate}
  \item The definition of $g_1\leq g_2$ is self-dual, \ie, $g_1\leq g_2$ in
  $\C$ $\iff$ $g_1\leq g_2$ in $\C^{op}$.

  \item A pair $(e,p)$ is an e-p pair from $X$ to $Y$ in $\C$ $\iff$ $(p,e)$ is
  an e-p pair from $X$ to $Y$ in $\C^{op}$.  In particular, the swap
  functor%
  \begin{tikzcd}
    S:\C_{ep}\arrow[r] &(\C^{op})_{ep},
  \end{tikzcd}
  which maps $(e,p)$ to
  $(p,e)$, is an isomorphism of categories.
  \end{enumerate}

\end{proposition}

\begin{proposition}[Basic facts]\label{thm:idem-basic}
  In any category $\C$, the following hold:
  \begin{enumerate}
  \item \label{item:idemp_order_poset} the relation $\leq$ is a
    partial order (on idempotents), \ie,
    $g_1\leq g_2\leq g_1\implies g_1=g_2$ and
    $g_1\leq g_2\leq g_3\implies g_1\leq g_3$. Moreover,
    $g\leq g\leq\id_X$, when $g$ is an idempotent on $X$.
  \item \label{item:e_o_p_idempotent} if $(e,p)$ is an e-p pair, then $g=e\circ p$ is
    an idempotent.

  \item\label{thm:idem-basic:sup}
    if $(g_i \mid i:I)$ is a family of idempotents on $X$ which is
    \emph{jointly mono},\ie,
    \begin{equation*}
     (\forall i.g_i\circ f=g_i\circ
    f')\implies f=f',
  \end{equation*}
  then its sup is $\id_X$. Similarly, when $(g_i \mid i:I)$ is \emph{jointly
    epi}, \ie,
  \begin{equation*}
   (\forall i.f\circ g_i=f'\circ g_i)\implies f=f',
 \end{equation*}
 then its sup is $\id_X$.

  \item\label{thm:idem-basic:mono} every arrow in $\C_{ep}$ is a
    mono. Furthermore, $(e,p)$ is an isomorphism in $\C_{ep}$ $\iff$
    $p$ is the inverse of $e$ in $\C$.
  \item\label{thm:idem-basic:split}
    if%
    \begin{tikzcd}
      (e_i,p_i):X_i\arrow[r, Rightarrow] & Y
    \end{tikzcd}
    is a splitting of $g_i$ (for $i=1,2$), then $g_1\leq g_2\iff$
    there exists (necessarily unique)%
    \begin{tikzcd}
      (e,p):X_1\arrow[r, Rightarrow] & X_2
    \end{tikzcd}
    such that $(e_1,p_1)=(e_2,p_2)\circ(e,p)$.
  \end{enumerate}
\end{proposition}
\begin{proof}
  The proofs for Items~\ref{item:idemp_order_poset}
  and~\ref{item:e_o_p_idempotent} are straightforward. For the other
  items, we have:

  \paragraph{Item \ref{thm:idem-basic:sup}.}
  Consider an idempotent $g$ of $X$ which is an upper-bound of
  $(g_i \mid i)$ (\ie, $\forall i.g_i\leq g$). Then,
  $\forall i.g_i\circ g=g_i$, which implies $g=\id_X$, because
  $(g_i \mid i)$ is jointly mono.

  \paragraph{Item \ref{thm:idem-basic:mono}.}
  To prove that every arrow%
  \begin{tikzcd}
    (e,p):X\arrow[r, Rightarrow] & Y
  \end{tikzcd}
  in $\C_{ep}$ is a mono it suffices to observe that in $\C$ every
  embedding $e$ is mono (\ie, $e\circ f_1=e\circ f_2\implies f_1=f_2$)
  and every projection $p$ is epi (the dual of mono).
  If $(e',p')$ is the inverse of $(e,p)$ in $\C_{ep}$, then $e'$ is
  the inverse of $e$ in $\C$, thus $p=p\circ e\circ e'=e'$, because
  $p\circ e=\id_X$.

  \paragraph{Item \ref{thm:idem-basic:split}.}
  Uniqueness of $(e,p)$ is immediate, because $(e_2,p_2)$ is a mono in
  $\C_{ep}$.  For existence, define $e=p_2\circ e_1$ and
  $p=p_1\circ e_2$. Then, from the assumption
  $e_1\circ p_1=g_1\leq g_2=e_2\circ p_2$ we derive:

  \begin{enumerate}
  \item $p\circ e=\id_{X_1}$ (\ie, $(e,p)$ is an e-p pair), because

    $(p_1\circ e_2)\circ(p_2\circ e_1)=$ by definition of $g_2$\\
    $p_1\circ g_2\circ e_1=$ by $(e_1,p_1)$ e-p pair\\
    $p_1\circ g_2\circ e_1\circ(p_1\circ e_1)=$ by definition of $g_1$\\
    $p_1\circ g_2\circ g_1\circ e_1=$ by $g_1\leq g_2$\\
    $p_1\circ g_1\circ e_1=$ by definition of $g_1$\\
    $p_1\circ (e_1\circ p_1)\circ e_1=$ by $(e_1,p_1)$ e-p pair\\
    $\id_{X_1}\circ\id_{X_1}=\id_{X_1}$

  \item $e_2\circ e=e_1$, because

    $e_2\circ(p_2\circ e_1)=$ by definition of $g_2$\\
    $g_2\circ e_1=$  by $(e_1,p_1)$ e-p pair\\
    $g_2\circ e_1\circ(p_1\circ e_1)=$ by definition of $g_1$\\
    $g_2\circ g_1\circ e_1=$ by $g_1\leq g_2$\\
    $g_1\circ e_1=$ by definition of $g_1$\\
    $(e_1\circ p_1)\circ e_1=e_1$ by $(e_1,p_1)$ e-p pair

  \item $p\circ p_2=p_1$, dual of the previous proof
  \end{enumerate}
\end{proof}

\begin{proposition}[$\omega$-colimits of embeddings]\label{thm:embed}
  Given an $\omega$-chain:%
  \begin{tikzcd}
    ((e_n,p_n):X_n\arrow[r, Rightarrow] & X_{n+1} \mid n:\omega)
  \end{tikzcd}
  in $\C_{ep}$ and a colimit cone $(\ul{f}_n:X_n\to \ul{X} \mid n)$ in
  $\C$ from the $\omega$-chain $(e_n:X_n\to X_{n+1} \mid n)$ to
  $\ul{X}$, there exists a unique cone
  \begin{tikzcd}
    ((\ul{f}_n,\ul{q}_n):X_n \arrow[r, Rightarrow] & \ul{X} \mid n)
  \end{tikzcd}
  in $\C_{ep}$ from the $\omega$-chain $((e_n,p_n) \mid n)$ to $\ul{X}$.
  Moreover, if $\ul{g}_n$ is the idempotent on $\ul{X}$ defined by the
  e-p pair $(\ul{f}_n,\ul{q}_n)$, then $(\ul{g}_n \mid n)$ is a jointly epi
  $\omega$-chain of idempotents on $\ul{X}$.
\end{proposition}

\begin{proof}
  First, we define the family $(h_{i,j}:X_i \to X_j \mid i,j:\omega)$
  of maps in $\C$ (by induction on $|j-i|$)
  \begin{itemize}
  \item $h_{i,i}$ is the identity on $X_i$
  \item $h_{i,j}=h_{i+1,j}\circ e_i$ when $i<j$, thus $h_{i,i+1}=e_i$
  \item $h_{i,j}=p_j\circ h_{i,j+1}$ when $i>j$, thus $h_{i+1,i}=p_i$
  \end{itemize}
  \noindent
  Second, we prove (by induction on $|j-i|$) the property
  $\forall i,j.h_{i+1,j}\circ e_i=h_{i,j}=p_j\circ h_{i,j+1}$.
  \begin{itemize}
  \item case $i=j$: immediate, since $h_{i+1,i}=p_i$ and
    $h_{i,i+1}=e_i$.
  \item case $i<j$: $h_{i+1,j}\circ e_i=h_{i,j}$ by definition of
    $h_{i,j}$.  For the other equality:
    \begin{enumerate}
    \item $h_{i,j}=$ by definition of $h_{i,j}$
    \item $h_{i+1,j}\circ e_i=$ by IH on $(i+1,j)$
    \item $p_j\circ h_{i+1,j+1}\circ e_i=$ by definition of $h_{i,j+1}$
    \item $p_j\circ h_{i,j+1}$.
    \end{enumerate}
  \item case $i>j$: $h_{i,j}=p_j\circ h_{i,j+1}$ by definition of
    $h_{i,j}$.  For the other equality:
    \begin{enumerate}
    \item $h_{i+1,j}\circ e_i=$ by definition of $h_{i+1,j}$
    \item $p_j\circ h_{i+1,j+1}\circ e_i=$ by IH on $(i,j+1)$
    \item $p_j\circ h_{i,j+1}=$ by definition of $h_{i,j}$
    \item $h_{i,j}$.
    \end{enumerate}
  \end{itemize}
  The property implies that $(h_{i,n} \mid i)$ is a cone in $\C$ from
  $(e_i \mid i)$ to $X_n$. Thus, there exists a unique
  $\ul{q}_n:\ul{X} \to X_n$ such that
  $\forall i.\ul{q}_n\circ \ul{f}_i=h_{i,n}$. In particular,
  $\ul{q}_n\circ \ul{f}_n=h_{n,n}=\id_{X_n}$, \ie,
  \begin{tikzcd}
    (\ul{f}_n,\ul{q}_n):\ul{X}\arrow[r, Rightarrow] & X_n.
  \end{tikzcd} 
  The property implies also that $\ul{q}_n=p_n\circ \ul{q}_{n+1}$,
  because $(\ul{f}_n \mid n)$ is jointly epi. Thus,
  $((\ul{f}_n,\ul{q}_n) \mid n)$ is a cone in $\C_{ep}$ from $\ul{X}$
  to $((e_n,p_n) \mid n)$.

     To prove uniqueness of $(\ul{q}_n \mid n)$, we use again that
     $(\ul{f}_n \mid n)$ is jointly epi and prove (by induction on
     $|j-i|$) that $\forall i,j.q'_j\circ \ul{f}_i=h_{i,j}$ when
     $(q'_n \mid n)$ is a cone from $\ul{X}$ to $(p_n \mid n)$ such
     that $\forall n.q'_n\circ \ul{f}_n=\id_{X_n}$.
  \begin{itemize}
  \item case $i=j$: immediate, since $q'_i\circ \ul{f}_i=\id_{X_i}=h_{i,i}$
    by assumption on $(q'_n \mid n)$ and definition of $h_{i,i}$.
  \item case $i<j$:
    \begin{enumerate}
    \item $q'_j\circ \ul{f}_i=$ by definition of $(\ul{f}_n \mid n)$
    \item $q'_j\circ \ul{f}_{i+1}\circ e_i=$ by IH on $(i+1,j)$
    \item $h_{i+1,j}\circ e_i=h_{i,j}$ by definition of $h_{i,j}$.
    \end{enumerate}
  \item case $i>j$:
    \begin{enumerate}
    \item $q'_j\circ \ul{f}_i=$ by assumption on $(q'_n \mid n)$
    \item $p_j\circ q'_{j+1}\circ \ul{f}_i=$ by IH on $(i,j+1)$
    \item $p_j\circ h_{i,j+1}=h_{i,j}$ by definition of $h_{i,j}$.
    \end{enumerate}
  \end{itemize}
  \noindent
  Consider the idempotents $\ul{g}_n=\ul{f}_n\circ \ul{q}_n$ on $\ul{X}$.
  From item~\ref{thm:idem-basic:split} of
  Proposition~\ref{thm:idem-basic}, it follows that
  $\forall n.\ul{g}_n\leq \ul{g}_{n+1}$.
  The family $(\ul{g}_n \mid n)$ is jointly epi, because $(\ul{f}_n \mid n)$ is
  jointly epi (as colimit cones are jointly epi) and each $\ul{q}_n$
  is (split) epi.
\end{proof}

  The following is the dual of Proposition~\ref{thm:embed}:
\begin{corollary}\label{thm:embed:dual}
  Given an $\omega$-chain%
  \begin{tikzcd}
    ((e_n,p_n):X_n \arrow[r, Rightarrow] & X_{n+1} \mid n)
  \end{tikzcd}
  in $\C_{ep}$ and a limit cone $(\ol{q}_n:\ol{X} \to X_n \mid n)$ in
  $\C$ from $\ol{X}$ to the $\omega^{op}$-chain
  $(p_n:X_{n+1} \to X_n \mid n)$, there exists a unique cone:
  \begin{equation*}
  \begin{tikzcd}
    ((\ol{f}_n,\ol{q}_n):X_n \arrow[r, Rightarrow] & \ol{X} \mid n)
  \end{tikzcd} 
\end{equation*}
  in $\C_{ep}$ from the $\omega$-chain $((e_n,p_n) \mid n)$ to $\ol{X}$.
  Moreover, if $\ol{g}_n$ is the idempotent on $\ol{X}$ defined by the
  e-p pair $(\ol{f}_n,\ol{q}_n)$, then $(\ol{g}_n \mid n)$ is a jointly mono
  $\omega$-chain of idempotents on $\ol{X}$.
\end{corollary}

The following result implies existence and uniqueness of a map
$\iota:\ul{X} \to \ol{X}$ in $\C$ such that
$\forall n.\iota\circ\ul{f}_n=\ol{f}_n$ and
$\forall n.\ul{q}_n=\ol{q}_n\circ\iota$, where%
\begin{tikzcd}
  ((\ul{f}_n,\ul{q}_n):X_n \arrow[r, Rightarrow] & \ul{X} \mid n)
\end{tikzcd}
and
\begin{tikzcd}
  ((\ol{f}_n,\ol{q}_n):X_n \arrow[r, Rightarrow] & \ol{X} \mid n)
\end{tikzcd}
are the cones in $\C_{ep}$ given by Proposition~\ref{thm:embed} and
Corollary~\ref{thm:embed:dual}.
In general, there is no reason for $\iota$ to be mono, epi or iso,
\eg: in $\Set$ the map $\iota$ is always mono, but it may fail to be
epi; in $\Set^{op}$ the converse holds; in $\Haus$ the map $\iota$ is
both mono and epi, but $\ul{X}$ may fail to be a sub-space of
$\ol{X}$.

\begin{theorem}\label{thm:idempotent}
  If $(g_n \mid n:\omega)$ is an $\omega$-chain of idempotents on $X$ in
  $\C$ such that every $g_n$ has a splitting, say%
  \begin{tikzcd}
   (f_n,q_n):X_n \arrow[r, Rightarrow] & X,
 \end{tikzcd}
 then there exists a unique $\omega$-chain%
 \begin{tikzcd}
   ((e_n,p_n):X_n \arrow[r, Rightarrow] & X_{n+1} \mid n:\omega)
 \end{tikzcd}
 in $\C_{ep}$ such that%
 \begin{tikzcd}
   ((f_n,q_n):X_n\arrow[r, Rightarrow] & X \mid n)
 \end{tikzcd}
 is a cone in $\C_{ep}$ from%
 \begin{tikzcd}
   ((e_n,p_n):X_n\arrow[r, Rightarrow] & X_{n+1} \mid n)
 \end{tikzcd}
 to $X$.  Moreover:
  \begin{enumerate}
  \item\label{thm:idempotent:colimit} If
    $(\ul{f}_n:X_n \to \ul{X} \mid n)$ is a colimit cone in $\C$ from
    the $\omega$-chain $(e_n:X_n \to X_{n+1} \mid n)$ to $\ul{X}$ and
    $\ul{\iota}:\ul{X}\to X$ is the unique map such that
    $\forall n.f_n=\ul{\iota}\circ\ul{f}_n$, then
    $\forall n.q_n\circ\ul{\iota} = \ul{q}_n$ (see
    Proposition~\ref{thm:embed} for $\ul{q}_n$).

  \item\label{thm:idempotent:limit} If
    $(\ol{q}_n:\ol{X}\to X_n \mid n)$ is a limit cone in $\C$ from
    $\ol{X}$ to the $\omega^{op}$-chain $(p_n:X_{n+1}\to X_n \mid n)$
    and $\ol{\iota}:X\to \ol{X}$ is the unique map such that
    $\forall n.q_n=\ol{q}_n\circ\ol{\iota}$, then
    $\forall n.\ol{\iota}\circ f_n = \ol{f}_n$ (see
    Corollary~\ref{thm:embed:dual} for $\ol{f}_n$).
  \end{enumerate}
\end{theorem}

\begin{proof}
  By Item~\ref{thm:idem-basic:split} of
  Proposition~\ref{thm:idem-basic}, we get that for each $n:\omega$
  there exists a unique e-p pair $(e_n,p_n)$ such that
  $(f_{n+1},q_{n+1})\circ(e_n,p_n)=(f_n,q_n)$, which implies that%
  \begin{tikzcd}
    ((f_n,q_n):X_n\arrow[r, Rightarrow] &  X \mid n)
  \end{tikzcd}
  is a cone in $\C_{ep}$ from the $\omega$-chain%
  \begin{tikzcd}
    ((e_n,p_n):X_n\arrow[r, Rightarrow] & X_{n+1} \mid n)
  \end{tikzcd}
  to $X$.

  \paragraph{Item \ref{thm:idempotent:colimit}.}
  We rely on the proof of Proposition~\ref{thm:embed}, where
  $\ul{q}_n$ is defined.
  Since $(\ul{f}_i \mid i)$ is jointly epi,
  $\forall j.q_j\circ\ul{\iota} = \ul{q}_j$ follows from
  $\forall i,j.q_j\circ\ul{\iota}\circ\ul{f}_i =
  \ul{q}_j\circ\ul{f}_i$, or equivalently, from
  $\forall i,j.q_j\circ f_i = h_{i,j}$, which we prove by induction on
  $|j-i|$:
  \begin{itemize}
  \item case $i=j$: immediate, since $q_i\circ f_i=\id_{X_i}=h_{i,i}$,
    because $(f_i,q_i)$ is a map in $\C_{ep}$ and by definition of
    $h_{i,i}$.
  \item case $i<j$:
    \begin{enumerate}
    \item $q_j\circ f_i=$ because
      $(f_{i+1},q_{i+1})\circ(e_i,p_i)=(f_i,q_i)$ in $\C_{ep}$
    \item $q_j\circ f_{i+1}\circ e_i=$ by IH on $(i+1,j)$
    \item $h_{i+1,j}\circ e_i=h_{i,j}$ by definition of $h_{i,j}$.
    \end{enumerate}
  \item case $i>j$:
    \begin{enumerate}
    \item $q_j\circ f_i=$  because
      $(f_{i+1},q_{i+1})\circ(e_i,p_i)=(f_i,q_i)$ in $\C_{ep}$
    \item $p_j\circ q_{j+1}\circ f_i=$ by IH on $(i,j+1)$
    \item $p_j\circ h_{i,j+1}=h_{i,j}$ by definition of $h_{i,j}$.
    \end{enumerate}
  \end{itemize}

  \paragraph{Item \ref{thm:idempotent:limit}.}
  By duality, since it is the dual of Item \ref{thm:idempotent:colimit}.
\end{proof}

\begin{proposition}\label{thm:iota:epi}
  Given two cones $(f_i:X_i\to X \mid i:I)$ and
  $(f'_i:X_i\to X' \mid i:I)$ and a map $\iota:X\to X'$ such that
  $\forall i.\iota\circ f_i=f'_i$, then $(f'_i \mid i)$ is jointly epi
  implies $\iota$ is epi.
\end{proposition}
\begin{proof}
  We have to prove that $h\circ\iota=h'\circ\iota$ implies $h=h'$.
  \begin{enumerate}
  \item $h\circ\iota=h'\circ\iota$ implies
  \item $\forall i.h\circ\iota\circ f_i=h'\circ\iota\circ f_i$ implies,
    by $\iota\circ f_i=f'_i$
  \item $\forall i.h\circ f'_i=h'\circ f'_i$ implies,
        by $(f'_i \mid i)$ jointly epi
  \item $h=h'$
  \end{enumerate}
\end{proof}
The following result is the dual of Proposition~\ref{thm:iota:epi}.
\begin{corollary}\label{thm:iota:mono}
  Given two cones $(q_i:X\to X_i \mid i:I)$ and
  $(q'_i:X'\to X_i \mid i:I)$ and a map $\iota:X\to X'$ such that
  $\forall i.q'_i\circ\iota=q_i$, then $(q_i \mid i)$ is jointly mono
  implies $\iota$ mono.
\end{corollary}

\subsection{Extended Metric Spaces versus Compact Hausdorff Spaces}
\label{sec:main-construction}

The following result requires to move between four categories (and we have
added also $\Set$) using four functors (where%
\begin{tikzcd}
  {} \arrow[r] & {}
\end{tikzcd}
denotes a faithful functor and%
\begin{tikzcd}
  {} \arrow[r, hook] & {}
\end{tikzcd}
an inclusion of a full sub-category):
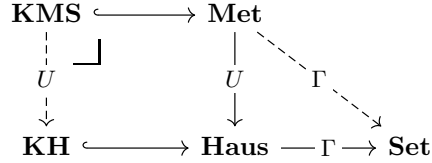
\begin{equation*}
  \begin{tikzcd}[row sep = huge, column sep = large]
    \KMS \arrow[r, hook] \arrow[d, "U" description, dashed]
    \arrow[dr, phantom, "\SEpbk", at start]
 & \MS \arrow[d, "U"  description] \arrow[dr,
    "\Gamma" description, dashed] &\\
    \KH \arrow[r, hook] & \Haus \arrow[r, "\Gamma" description] & \Set
  \end{tikzcd}
\end{equation*}
All categories in the diagram have:
\begin{itemize}
\item finite limits and finite sums (Theorem~\ref{thm-cat-KMS});
\item enough points, \ie, the global section functors\footnote{Given a
  category $\C$ with a terminal object $1$, the global section functor
  from $\C$ to $\Set$ is given by $\C(1,-)$.}  $\Gamma$ into $\Set$
  are faithful;
\item splittings of idempotents;
\end{itemize}
and all functors in the diagrams preserve finite limits and finite
sums. Moreover, $\Haus$ has all small limits and small colimits
(Theorem~\ref{thm-cat-KH}), where limits are computed as in $\Top$,
and limits (computed in $\Haus$) of diagrams in $\KH$ are in $\KH$.
Also $\MS$ has all small limits and small colimits
(Theorem~\ref{thm-cat-KMS}), but the forgetful functor $U$ from $\MS$
to $\Haus$ may not preserve all these (co)limits.

In applications, we start from a metric space $\State$, then identify
an $\omega$-chain $(g_n \mid n)$ of idempotents on $\State$ in $\MS$, and
by applying Theorem~\ref{thm:idempotent} in $\Haus$ we get a map
$\ol{\iota}:\State\to\ol{\State}$ in $\Haus$.
The theorems below provide sufficient conditions to ensure that
$\ol{\State}$ is compact, $\ol{\iota}$ is mono and epi, and, above
all, that the complete lattice $\cL(\ol{\State})$ is
$\omega$-continuous and the monotonic map $\cL(\ol{\iota})$ is in
$\mathbb{A}_{RS}(\State,\ol{\State})$.
These properties can be proved for maps $\iota:\State\to\ol{\State}$
that are not necessarily obtained through
Theorem~\ref{thm:idempotent}, and the theorems below capture this
greater generality.

If we start from an $\omega$-chain $(g_n \mid n)$ of idempotents on
$\State$ in $\MS$, then Theorem~\ref{thm:idempotent} provides
candidates for $(p_n \mid n)$ and $(q_n \mid n)$ in the following theorem,
because $\MS$ has splittings of idempotents.
\begin{theorem}\label{thm:KMS}
  If $(p_n:\State_{n+1}\to \State_n \mid n)$ is an $\omega^{op}$-chain
  in $\KMS$, $(q_n \mid n)$ is a cone from $\State$ to $(p_n \mid n)$
  in $\MS$, $(\ol{q}_n:\ol{\State}\to \State_n \mid n)$ is a limit
  cone from $\ol{\State}$ to $(p_n \mid n)$ in $\Haus$, and
  $\iota:\State\to \ol{\State}$ is the unique map in $\Haus$ such that
  $q_n=\ol{q}_n\circ \iota$, then:
\begin{enumerate}
\item $\ol{\State}$ is compact and has a countable base, thus
  $\cL(\ol{\State})$ is $\omega$-continuous.
\item The monotonic map $\cL(\iota)$ is in
  $\mathbb{A}_{RS}(\State,\ol{\State})$.
\end{enumerate}
\end{theorem}

\begin{proof}
  The Hausdorff space $\ol{\State}$ can be identified with the set
  $\{s:\prod_n\absn{\State_n} \mid \forall n. s_n=p_n(s_{n+1})\}$,
  equipped with the coarsest topology $\OS(\ol{\State})$ making the
  maps $\ol{q}_n(s)=s_n$ continuous, \ie, the topology generated by
  the sub-base $[O]_n=\{s:\absn{\ol{\State}} \mid \ol{q}_n(s)\in O\}$,
  where $O$ is an open set in $\State_n$.

By Theorem~\ref{thm-cat-KH}, the limit $\ol{\State}$ is in $\KH$,
because it is the limit (in $\Haus$) of a diagram in $\KH$.
  The topology on $\State_n$ has a countable base $\tau^b_n$, because
  $\State_n$ is in $\KMS$.  Thus, the topology on $\ol{\State}$ has a
  countable sub-base too, namely the set of open subsets of the form
  $[B]_n$ with $B\in\tau^b_n$.

  By Theorem~\ref{thm-top-S}, the complete lattice $\cL(X)$ is in
  $\CL$ and the topologies $\tau_S(X)$ and $\tau_U(X)$ coincide, when
  $X:\KH$, as in the case of $\State_n$ and $\ol{\State}$.  Therefore,
  $\tau_S(X)$ is generated by
  $\uparrow O\defeq\{K:\CS(X) \mid K\subseteq O\}$ with $O:\OS(X)$,
  and the way-below relation is given by
  $K_1\ll K_2\iff K_2\subseteq \interiorOf{K_1}$, where
  $\interiorOf{K_1}$ is the interior of $K_1$.

By Theorem~\ref{thm-cL2}, the continuous lattice $\cL(\ol{\State})$ is
(isomorphic to) the limit of the $\omega^{op}$-chain of right adjoints
$(\cL(p_n) \mid n)$ in $\CL$ (and in $\Po_A$). To be more precise, the
isomorphism is $K\mapsto(\ol{q}_n(K) \mid n)$ and
$(K_n \mid n)\mapsto\bigcap_n \ol{q}_n^*(K_n)$.

  The sub-base of $\OS(\ol{\State})$ given above, \ie,
  the set of $[O]_n=\{s:\absn{\ol{\State}} \mid \ol{q}_n(s)\in O\}$, where
  $O$ is an open set in $\State_n$, is actually a base, because
  $[O]_n=[p_n^*(O)]_{n+1}$.
  Therefore, every $O:\OS(\ol{\State})$ is of the form
  $\bigcup_{i:I}[O_i]_{n_i}$ with $O_i:\OS(\State_{n_i})$ for $i:I$.
  Since $\ol{\State}$ is compact, also $K:\CS(\ol{\State})$ is
  compact, and $K\subseteq O$ implies
  $K\subseteq\bigcup_{i:J}[O_i]_{n_i}$ for some $J\subseteq_f I$.
  In particular, $O\supseteq\bigcup_{i:J}[O_i]_{n_i}=[O_J]_{n_J}$,
  where $n_J=\sup_{i:J} n_i$ and $O_J:\OS(\State_{n_J})$ is the union
  for $i:J$ of the $O_i$ moved from $\State_{n_i}$ to $\State_{n_J}$,
  therefore
  \begin{equation}
    \label{key-prop}
    \forall K:\CS(\ol{\State}).\forall O:\OS(\ol{\State}).
  K\subseteq O\iff\exists n.\exists
  O_n:\OS(\State_n).\ol{q}_n(K)\subseteq O_n\land [O_n]_n\subseteq O.
\end{equation}
\noindent
Using the above property, we have:
\begin{itemize}
\item $\cL(\iota):\mathbb{A}_{RS}(\State,\ol{\State})$ means $\forall
  C:\CS(\State).\forall O:\OS(\ol{\State}).\cl{\iota(C)}\subseteq
  O\implies\exists\delta.\cl{\iota(C_\delta)}\subseteq O$.
\item By property~\ref{key-prop}, this is implied by $\forall
  C:\CS(\State).\forall n.\forall O:\OS(\State_n).\cl{\iota(C)}\subseteq
  [O]_n\implies\exists\delta.\cl{\iota(C_\delta)}\subseteq [O]_n$.
\item By $q_n=\ol{q}_n\circ \iota$ and the definition of $[O]_n$, this is
  equivalent to:
  \begin{equation*}
  \forall C:\CS(\State).\forall n.\forall
  O:\OS(\State_n).\cl{q_n(C)}\subseteq
  O\implies\exists\delta.\cl{q_n(C_\delta)}\subseteq O.
  \end{equation*}

\item In general, $q:\MS(\State,\State')$ implies $\forall\delta>0.\forall
  C:\CS(\State).\cl{q(C_\delta)}\subseteq\cl{q(C)}_\delta$,
  and\\ $\State':\KMS$ implies $\forall K:\CS(\State').\forall
  O:\OS(\State'). K\subseteq O\implies\exists\delta.K_\delta\subseteq
  O$.

\item Since $q_n:\MS(\State,\State_n)$ and $\State_n:\KMS$,
  for any $C:\CS(\State)$ and $O:\OS(\State_n)$\\
  $\cl{q_n(C)}\subseteq
  O \implies\exists\delta.\cl{q_n(C)}_\delta\subseteq O
  \implies\exists\delta.\cl{q_n(C_\delta)}\subseteq O$.
\end{itemize}
\end{proof}

If we start from an $\omega$-chain $(g_n \mid n)$ of idempotents on
$\State$ in $\MS$, rather than in $\Haus$, then $((f_n,q_n) \mid n)$
and $((e_n,p_n) \mid n)$ in the following theorem consist of short
maps.
Moreover, families of maps that are jointly mono in $\MS$ are also
jointly mono in $\Haus$, because these categories have enough points.
Finally, if an $\omega$-chain $(g_n \mid n)$ of idempotents on $\State$ (in the
category $\C$) is jointly mono (in $\C$), then the identity on $\State$ is
the sup of the $\omega$-chain in the poset of idempotents on $\State$.
\begin{theorem}\label{thm:idem}
  Given an $\omega$-chain $(g_n \mid n)$ of idempotents on $\State$ in
  $\Haus$, which is jointly mono, consider:
\begin{itemize}
\item a splitting
  \begin{tikzcd}[column sep = large]
    \State \arrow[r, "q_n" description] &\State_n \arrow[r, hook,
    "f_n" description] &\State
  \end{tikzcd}of $g_n$;
\item the unique $\omega$-chain
  \begin{tikzcd}
    ((e_n,p_n):\State_n\arrow[r, Rightarrow] &
  \State_{n+1} \mid n)
\end{tikzcd}
in $\Haus_{ep}$ such that
\begin{tikzcd}
  ((f_n,q_n):\State_n\arrow[r, Rightarrow] & \State \mid n)
\end{tikzcd}
is a cone in $\Haus_{ep}$ from
\begin{tikzcd}
  ((e_n,p_n):\State_n\arrow[r , Rightarrow] & \State_{n+1} \mid n)
\end{tikzcd}
to $\State$ (see Theorem~\ref{thm:idempotent});
\item the limit cone $(\ol{q}_n:\ol{\State}\to \State_n \mid n)$ in
  $\Haus$ from $\ol{\State}$ to the $\omega^{op}$-chain
  $(p_n:\State_{n+1}\to \State_n \mid n)$ (see
  Corollary~\ref{thm:embed:dual});
\item the unique map $\iota:\State\to \ol{\State}$ in $\Haus$ such
  that $\forall n.q_n=\cl{q}_n\circ \iota$.
\end{itemize}
\noindent
Then, $\iota$ is both mono and epi in $\Haus$.
\end{theorem}

\begin{proof}
  If $(g_n \mid n)$ is jointly mono, then also $(q_n \mid n)$ is jointly mono.
  Therefore, $\iota$ is mono by Corollary~\ref{thm:iota:mono}.

  If
  \begin{tikzcd}
    ((\ol{f}_n,\ol{q}_n):\State_n\arrow[r, Rightarrow] & \ol{\State} \mid n)
  \end{tikzcd}
  is the unique cone in $\Haus_{ep}$ given by
  Corollary~\ref{thm:embed:dual}, then
  $\forall n.\iota\circ f_n = \ol{f}_n$, by
  Theorem~\ref{thm:idempotent}.  Therefore, to prove that $\iota$ is
  epi it suffices, by Proposition~\ref{thm:iota:epi}, to prove that
  $(\ol{f}_n \mid n)$ is jointly epi in $\Haus$.
  This amounts to proving that the union of the images of the maps in
  $(\ol{f}_n \mid n)$ is dense in $\ol{\State}$.  To do this we use
  the base of $\OS(\ol{\State})$ as in the proof of
  Theorem~\ref{thm:KMS}.
  For every $s:\ol{\State}$ and $[O]_n$ in the base (\ie,
  $O:\OS(\State_n)$) such that $s:[O]_n$ (\ie, $\ul{q}_n(s):O$), we
  have to give an $s_n:\State_n$ such that $\ul{f}_n(s_n):[O]_n$.  It
  suffice to take $s_n=\ul{q}_n(s)$, since
  $s_n=\ul{q}_n(\ul{f}_n(s_n))$.
\end{proof}


\section{Examples}\label{sec:examples}

In this section, we consider examples of Banach spaces $\State$,
demonstrating how to apply the results of Section~\ref{sec:main} to
define a specific $\cl{\State}$ in $\KH$, and in which cases
$\cl{\State}$ is a compactification of $\State$ (a summary is given at
the end of this section, see Table~\ref{table:summary_of_examples}).
In Section~\ref{sec:precision}, we will study loss of precision when
going from $\State$ to $\cl{\State}$. All examples considered in this
section are \emph{sequence spaces}. Hence, we recall some general
definitions and fix notation.

\begin{definition}[Uniform notation]\label{def-ssp-cub}
  We write $\Real$ for the standard Banach space on the reals, and
  also for the underlying vector space, metric space, topological
  space, and set.

\begin{itemize}
\item Given a set $I$, we write $\Real^I$ for the product of $I$
  copies of the set $\Real$, which is also the carrier of the product
  of $I$ copies of $\Real$ in the categories of the vector spaces,
  extended metric spaces and Hausdorff spaces.
\item
Given a real number $p$ in the interval $[1,\infty)$, we write
$\norm{-}_{I,p}$ for the map from $\Real^I$ to $[0,\infty]$ given by
$\norm{x}_{I,p}\defeq(\sum_{i:I}\absn{x_i}^p)^{1/p}$, and it is
extended to $p=\infty$ by defining
$\norm{x}_{I,\infty}\defeq\sup_{i:I}\absn{x_i}$.
Since $I$ is determined by $x$, we drop the subscript $I$ and write
$\norm{x}_p$.

\item We write $\ssp{I,p}$ for the Banach space with carrier the
  sub-space $\{x:\Real^I \mid \norm{x}_p<\infty\}$ of (the vector
  space) $\Real^I$ and norm $\norm{-}_{I,p}$. We write $\cub{I,p}$ for
  the closed unit ball in $\ssp{I,p}$, whose elements are those $x$
  such that $\norm{x}_p\leq 1$.
The subset $\cub{I,p}$ inherits from $\ssp{I,p}$ the metric space
structure.

\item
If $I\subseteq J$, then $\ssp{I,p}$ is isomorphic (in the category of
Banach spaces and short linear maps) to the sub-space of $\ssp{J,p}$
with carrier $\{x \mid \forall j:J \setminus I.x_j=0\}$, and $\cub{I,p}$ is a sub-space of $\cub{J,p}$ (modulo the isomorphism).
\end{itemize}
We consider only countable $I$, specifically, either $\omega$ or a
natural number $m$.  We write $\ssp{p}$ for $\ssp{\omega,p}$ and
$\ssp{*,p}$ for the (normed vector) sub-space of $\ssp{p}$ with
carrier $\{x \mid \exists n.\forall i>n.x_i=0\}$.  Usually
$\ssp{*,\infty}$ is denoted $c_{00}$. We write
$\cub{p}$ for $\cub{\omega,p}$, and $\cub{*,p}$ for
$\cub{p}\cap\ssp{*,p}$.
Note that $\ssp{0,p}$ is trivial and $\ssp{1,p}=\Real$ for every $p$.
\end{definition}

In the sequel, we use the following characterization of limits in
$\Top$ and general properties of limits and colimits (valid in any
category).

\begin{proposition}
  \label{thm:lim-top}

  Given a small diagram $D:I\to \Top$, a limit cone
  $(\pi_i:(X,\tau)\to D_i \mid i:I)$ in $\Top$ is obtained by taking a
  limit cone $(\pi_i:X\to U(D_i) \mid i:I)$ of $U\circ D:I\to \Set$ in
  $\Set$, and by defining $\tau$ as the coarsest topology on $X$
  making the maps $\pi_i:(X,\tau)\to D_i$ continuous.
\end{proposition}

\begin{proposition}[Limits commute with Limits]
  \label{thm:lim-commute}
  Given an $I \times J$-diagram $D:I \times J\to \C$ in a category
  $\C$ (with the relevant limits), if for each $i:I$,
  $(p^i_j:X_i\to D_{i,j} \mid j:J)$ is a limit cone for the
  $J$-diagram $D(i,-):J\to \C$, then the family $(X_i \mid i:I)$
  extends \emph{canonically} to an $I$-diagram $X:I\to \C$, namely,
  for $f:i\to i'$ in $I$, the map $X_f:X_i\to X_{i'}$ is the unique
  map in $\C$ such that for all $j:J$, the following diagram
  commutes:%
  \begin{equation*}
  \begin{tikzcd}[row sep = large, column sep = large]
    X_i \arrow[r, dashed, "X_f" description] \arrow[d, "p^i_j" description] & X_{i'} \arrow[d,
    "p^{i'}_j" description]\\
    D_{i,j} \arrow[r, "D_{f,j}" description] & D_{i',j}
  \end{tikzcd}
  \end{equation*}
  Moreover, if $(p_i:x\to X_i \mid i:I)$ is a limit cone for
  $X:I\to \C$, then $(p^i_j\circ p_i:x\to D_{i,j} \mid i:I,j:J)$ is a
  limit cone for $D$. Since one can exchange the role of $I$ and $J$,
  there are two alternative ways of computing limits of
  $I \times J$-diagrams, which necessarily produce canonically
  isomorphic results.
\end{proposition}

\begin{proposition}[Colimits of cofinal diagrams]
  \label{thm:cofin-colim}
  Given an $\omega$-diagram $D:\omega\to \C$ in a category $\C$ (with
  the relevant colimits), if $(f_n:D_n\to X \mid n:\omega)$ is a
  colimit cone for $D$ and $h:\omega\to \omega$ is a strictly
  increasing map, then $(f_{h(n)}:D_{h(n)}\to X \mid n:\omega)$ is a
  colimit cone for the $\omega$-diagram $D\circ h:\omega\to \C$.
\end{proposition}

\subsection{Banach space $\Real$}
\label{subsec:examples_real_line}

Consider the metric space $\Real$ with distance $d(x,y) = \absn{x -
y}$ and the $\omega$-chain $(r_n \mid n:\omega)$ such that
\begin{equation}
  \label{eq:def_r_n}
  r_n(x)\defeq\left\{
    \begin{array}{ll}
      n,  & \text{ if } n < x, \\
      x,  & \text{ if } \absn{x} \leq n, \\
      -n, & \text{ if }  x < -n. \\
    \end{array}
    \right.
\end{equation}
Each $r_n$ is idempotent and short, because:
\begin{equation*}
 d(r_n(x),r_n(y))=\left\{
    \begin{array}{c|c|c||l}
      x<-n&-n\leq x\leq n&n<x\\\hline \hline
      0&n+x&2n&y<-n\\\hline
      n+y&d(x,y)&n-y&-n\leq y\leq n\\\hline
      2n&n-x&0&n<y\\
      \end{array}\right\}\leq d(x,y) .
  \end{equation*}
The image of $r_n$ is the compact sub-space $\Real_n\defeq [-n,n]$,
and the union $\State_*$ of the sub-spaces $\Real_n$ is $\Real$.

Let%
\begin{tikzcd}[column sep = large]
  \Real \arrow[r, "q_n" description]&\Real_n \arrow[r, "f_n"
  description] &\Real
\end{tikzcd}be the splitting of $r_n$ through $\Real_n$ and
$p_n=q_n\circ f_{n+1}:\Real_{n+1}\to \Real_n$. Let $(\cl{q}_n \mid n)$
be the limit cone from $\cl{\State}$ to the $\omega^{op}$-chain
$(p_n \mid n)$ in $\Haus$. Then, by Theorem~\ref{thm:KMS},
$\cl{\State}$ is compact, and by Theorem~\ref{thm:idem}, the map
$\iota:\Real\to \cl{\State}$ is both epi and mono in $\Haus$.

We show that $(\cl{q}_n \mid n)$ is isomorphic to the cone $(\hat{q}_n \mid n)$
from $\cl{\Real}=[-\infty,+\infty]$ (the two-point
compactification of $\Real$) to $(p_n \mid n)$, where $\hat{q}_n$ is the
extension of $q_n$ to $\cl{\Real}$ mapping $-\infty$ to $-n$
and $+\infty$ to $+n$.
Let $\phi:\cl{\Real}\to \cl{\State}$ be the unique map such that
$\forall n.\hat{q}_n=\cl{q}_n\circ\phi$, namely
$\phi(x)\defeq(\hat{q}_n(x) \mid n)$.
The map $\phi$ is a bijection (in $\Set$), since the elements
$(s_n \mid n)$ in $\cl{\State}$ satisfy one of the following disjoint
properties:
\begin{itemize}
\item $\forall n.s_n=-n$, \ie, $(s_n \mid n)=\phi(-\infty)$;
\item $(s_n \mid n)$ is eventually constant. This happens when
  $\absn{s_m}<m$ for some $m:\omega$.  In this case
  $(s_n \mid n)=\phi(s_m)$;
\item $\forall n.s_n=+n$, \ie, $(s_n \mid n)=\phi(+\infty)$.
\end{itemize}
Therefore, $(\hat{q}_n \mid n)$ is a limit cone from $\cl{\Real}$ to
$(p_n \mid n)$ in $\Set$.
To prove that $\phi$ is an isomorphism in $\Top$, it suffices to show
that the topology on $\cl{\Real}$ is the coarsest topology making the
maps $\hat{q}_n$ continuous~(Proposition~\ref{thm:lim-top}). Since a
base for the topology on $\cl{\Real}$ consists of the subsets of the
form $[-\infty,x)$, $(x,y)$ and $(y,+\infty]$ for $x,y:\Real$, it
suffices to show that every element in the base is of the form
$\inv{\hat{q}_n}(O)$ for some $n:\omega$ and open subset
$O:\OS(\Real_n)$.  This is immediate by taking $n$ such that
$\absn{x},\absn{y}<n$, and taking $O$ of the form $[-n,x)$, $(x,y)$
and $(y,+n]$, respectively.

\subsection{Banach spaces $\ssp{m,\infty}$ for $1<m$}
\label{subsec:examples_ell_m_infty}

Fix a natural number $m>1$, consider the metric space
$\State=\ssp{m,\infty}$ with distance
$d_\infty(x,y)=\max_{i:m}d(x_i,y_i)$, which coincides with the finite
product $\Real^m$ in $\MS$, and the $\omega$-chain
$(g_n \mid n:\omega)$ defined by:
\begin{equation} \label{eq:def_g_m_n} \forall x:\Real^m. \
g_n(x) \defeq (r_n(x_i) \mid i:m), \end{equation}
where $r_n$ is as defined in~\eqref{eq:def_r_n}. Since $g_n$ is
defined pointwise, it is idempotent and short
by \emph{inheritance}, since $d_\infty(g_n(x),g_n(y))= \max_{i:m}
d(r_n(x_i),r_n(y_i))\leq \max_{i:m} d(x_i,y_i)= d_\infty(x,y)$.
The image of $g_n$ is the compact sub-space $\State_n\defeq\Real_n^m$
and, once again, the union $\State_*$ of the sub-spaces $\State_n$ is
$\State$.

From Section~\ref{subsec:examples_real_line} and
Proposition~\ref{thm:lim-commute} we have that $\cl{\State}$ is
isomorphic to $\cl{\Real}^m$ in $\KH$.
In fact, $\KH$ has all small limits. Thus, we can take $I=m$ and
$J=\omega^{op}$, and consider the $I \times J$-diagram
$D:I \times J\to \KH$ such that $D(i,n)=\Real_n$ and $D(i,n+1\to n)$
is the map $p_n:\Real_{n+1}\to \Real_n$ defined in
Section~\ref{subsec:examples_real_line}. The limit $\cl{\State}$ is
obtained by first computing the limits $\Real_n^m$ of $I$-diagrams
$D(-,n)$ and then the $J$-limit, while $\cl{\Real}^m$ is obtained by
first computing the limits $\cl{\Real}$ of the $J$-diagrams $D(i,-)$
and then the $I$-limit.

\subsection{Banach spaces $\ssp{m,p}$ for $1<m$ and $1\leq p<\infty$}
\label{subsec:examples_ell_m_p}

This is a modification of Section~\ref{subsec:examples_ell_m_infty},
where we consider the metric space $\State=\ssp{m,p}$ that has the
carrier of $\ssp{m,\infty}=\Real^m$, but with distance
\begin{equation*}
d_p(x,y)\defeq\left( \sum_{i:m} d(x_i,y_i)^p \right)^{1/p}\geq d_\infty(x,y).
\end{equation*}
We take the same $g_n$ used for $\ssp{m,\infty}$, as defined
in~\eqref{eq:def_g_m_n}. Clearly $g_n$ is idempotent, since this
property does not depend on the distance, and is short also with respect to~$d_p$
(again by inheritance), since
\begin{equation*}
  d_p(g_n(x),g_n(y))= \left(\sum_{i:m} d(r_n(x_i),r_n(y_i))^p \right)^{1/p} \leq \left( \sum_{i:m}
  d(x_i,y_i)^p \right)^{1/p} = d_p(x,y).
\end{equation*}
Therefore, the metric space $\State_n$ has the same carrier of
$\Real_n^m$ and distance $d_p$.  Since $d_p$ and $d_\infty$ induce the
same topology on $\Real_n^m$, we have that $\cl{\State}$ for
$\ssp{m,p}$ and for $\ssp{m,\infty}$ are equal and isomorphic to
$\cl{\Real}^m$ in $\KH$.


\subsection{Banach space $\ssp{\infty}$}
\label{subsec:examples_ell_infty}

Consider the metric space $\State=\ssp{\infty}$ with distance
$d_\infty(x,y)\defeq\sup_{i:\omega}d(x_i,y_i)$, and the $\omega$-chain
$(g_n \mid n:\omega)$ of maps on $\ssp{\infty}$ defined by
\begin{equation}
  \label{eq:g_n:ell_infty_ell_p}
\forall x:\ssp{\infty}. \ g_n(x) \defeq (r_n(x_i) \mid i:n)\cdot
0^\omega .
\end{equation}
Since $g_n$ is defined pointwise, it is idempotent and short
by \emph{inheritance}, \eg,
$$d_\infty(g_n(x),g_n(y))=
\sup_{i:n} d(r_n(x_i),r_n(y_i))\leq \sup_{i:n} d(x_i,y_i)\leq d_\infty(x,y).$$
The image of $g_n$ is the compact sub-space
$\State_n={\ssp{\infty}}_n\defeq \Real_n^n\times [0]^\omega$, which is
isomorphic to the finite product $\Real_n^n$ in $\MS$, and the union
$\State_*$ of the sub-spaces $\State_n$ is the sub-space
$\ssp{*,\infty}$ of $\omega$-sequences eventually equal to $0$, which
is not dense in $\ssp{\infty}$ and its closure is the sub-space $c_0$
of $\omega$-sequences converging to $0$.

From Section~\ref{subsec:examples_real_line},
Proposition~\ref{thm:lim-commute}, and the dual of
Proposition~\ref{thm:cofin-colim}, we have that $\cl{\State}$ for
$\ssp{\infty}$, which we denote with $\kssp{\infty}$, is isomorphic to
$\cl{\Real}^\omega$ in $\KH$.
In fact, take $I=\Nat$, $J=\omega^{op}$, and consider the
$I\X J$-diagram $D:I \times J\to \KH$ such that $D(i,n)=\Real_n$ if
$i<n$, else $[0]$, and $D(i,n+1\to n)= p_n: \Real_{n+1}\to \Real_n$ if
$i<n$, else the unique map from $D(i,n+1)$ to $[0]$.
The limit $\cl{\State}$ is obtained by first computing the limits of
$I$-diagrams $D(-,n)$, which are isomorphic to $\Real_n^n$, and then
the $J$-limit, while $\cl{\Real}^\omega$ is obtained by first
computing the limits $\cl{\Real}$ of the $J$-diagrams $D(i,-)$
and then the $I$-limit.

Note that the map $\iota:\ssp{\infty}\to \kssp{\infty}$ given by
Theorem~\ref{thm:idem} is mono and epi in $\Haus$ (the subset
$\ssp{*,\infty}$ is not dense in $\ssp{\infty}$, but it is dense in
$\kssp{\infty}$).

\subsection{Banach spaces $\ssp{p}$ for $1\leq p<\infty$}
\label{subsec:examples_ell_p}

Consider the metric spaces $\State=\ssp{p}$ with distance
$d_p(x,y)\defeq\left(\sum_{i:\omega}d(x_i,y_i)^p\right)^{1/p}\geq
d_\infty(x,y)$. The carrier $\absn{\ssp{p}}$ of $\ssp{p}$ satisfies
the following strict inclusions
\begin{equation*}
  \absn{\ssp{*,\infty}} \subset \absn{\ssp{p}} \subset \absn{c_0}
  \subset \ssp{\infty}.
\end{equation*}
We can consider the restrictions to $\ssp{p}$ of the
idempotents $g_n$ defined in~\eqref{eq:g_n:ell_infty_ell_p}.
It is straightforward to prove that $g_n$ is short and idempotent on
$\ssp{p}$, its image $\State_n={\ssp{p}}_n$ is isomorphic to the
metric space with carrier $\Real_n^n$ and distance $d_p$, and the
union $\State_*$ of the sub-spaces $\State_n$ is the dense sub-space
$\ssp{*,p}$ of $\ssp{p}$.

By analogy with Section~\ref{subsec:examples_ell_infty}, we have that
the map $\iota:\State\to \cl{\State}$ is mono and epi in $\Haus$, and
$\cl{\State}$ for $\ssp{p}$ is isomorphic to $\cl{\Real}^\omega$ in
$\KH$. In particular, $\cl{\State}$ is independent of $p$.
In summary, the relations between $\ssp{p}$ and $\ssp{q}$, for
$1\leq p<q\leq\infty$, are:

\begin{itemize}
\item The carrier of $\ssp{p}$ is a proper subset of the carrier of
  $\ssp{q}$, and the inclusion of $\ssp{p}$ into $\ssp{q}$ is a short
  map in $\MS$, since $\forall x,y:\ssp{p}.d_q(x,y)\leq d_p(x,y)$.

\item The compact metric spaces ${\ssp{p}}_n$ and ${\ssp{q}}_n$ have the same
      carrier, but different distances, and the inclusion of
      ${\ssp{p}}_n$ into ${\ssp{q}}_n$ is a short map in $\KMS$.
    \item As topological spaces, ${\ssp{p}}_n$ and ${\ssp{q}}_n$ are
      equal. Thus, $\kssp{p}=\kssp{q}$.

\end{itemize}

\subsection{Unit ball $\cub{\infty}$}
\label{subsec:examples_ell_infty_unit_ball}

Let us now consider the unit ball $\cub{\infty}$ in the metric space
$\ssp{\infty}$.
As a metric space, $\cub{\infty}$ coincides with the infinite product
$\Real_1^\omega$ in $\MS$, and the idempotents $g_n$ defined is
Section~\ref{subsec:examples_ell_infty} restrict to idempotents on
$\cub{\infty}$.
The image of $g_n$ (restricted to $\cub{\infty}$) is the compact sub-space
$\State_n={\cub{\infty}}_n\defeq\Real_1^n\times [0]^\omega$, which is
isomorphic to the finite product $\Real_1^n$ in $\MS$, and coincides
with closed unit ball $\cub{n,\infty}$ in $\ssp{n,\infty}=\Real^n$.
The union $\State_*$ of the sub-spaces $\State_n$ is the sub-space
$\cub{*,\infty}$, and the closure of $\State_*$ is the sub-space
$\cub{\infty}\bigcap c_0$.

Similar to Section~\ref{subsec:examples_ell_infty}, we can prove that
$\cl{\State}$ for $\cub{\infty}$, which we denote with
$\kcub{\infty}$, is isomorphic to product $\Real_1^\omega$ in $\KH$.
More precisely, take $I=\Nat$, $J=\omega^{op}$, and consider the
$I \times J$-diagram $D:I \times J\to \KH$ such that $D(i,n)=\Real_1$
if $i<n$ else $[0]$, and $D(i,n+1\to n)$ is the identity on $\Real_1$
if $i<n$, else the unique map from $D(i,n+1)$ to $[0]$.

As in the case of $\ssp{\infty}$, the map
$\iota:\cub{\infty}\to \kcub{\infty}$ given by
Theorem~\ref{thm:idem} is mono and epi in $\Haus$.
Moreover, as a set-theoretic map, $\iota$ is a bijection. In fact, the
metric space $\cub{\infty}$ is (equal to) the product $\Real_1^\omega$
in $\MS$ and the topological space $\kcub{\infty}$ is (isomorphic to)
the product $\Real_1^\omega$ in $\KH$. As such, without loss of
generality, we assume that the map $\iota$ is an identity map.
%

\subsection{Unit ball $\cub{p}$ for $1 \leq p < \infty$}
\label{subsec:examples_ell_p_unit_ball}

Let us now consider the unit ball $\cub{p}$ in the metric space
$\ssp{p}$. One can proceed in analogy with
Section~\ref{subsec:examples_ell_infty_unit_ball}. In particular,
${\cub{p}}_n$ is isomorphic to the closed unit ball $\cub{n,p}$ in
$\ssp{n,p}$, the map $\iota:\cub{p}\to \kcub{p}$ given by
Theorem~\ref{thm:idem} is mono and epi in $\Haus$.

Moreover, as a set-theoretic map, $\iota$ is a bijection. The map
$\iota$ is clearly injective. So, it suffices to prove that it is
surjective. Let us consider an element $(x_n \mid n)$ in
$\kcub{p}$. Each $x_n$ may differ from $x_{n+1}$ only in the $n$-th
component, namely $x_{n,n}=0$, while $x_{n+1,n}:\Real_1$. Consider
$y:\Real_1^\omega$ defined by $y_n\defeq x_{n+1,n}$ for $n:\omega$. We
have $\norm{y}_p=\lim_{n \to \infty}\norm{x_n}_p \leq 1$, which
entails that $y:\cub{p}$. Furthermore,
\begin{equation*}
  \forall i,n : \omega. \quad x_{n,i}=
  \left\{
    \begin{array}{ll}
      y_i, & \text{ if } i < n,\\
      0, & \text{otherwise}.
    \end{array}
  \right.
\end{equation*}
Hence, $\iota(y) = (x_n \mid n)$, and $\iota$ is surjective.

The relations established in Section~\ref{subsec:examples_ell_p} imply
the following relations between $\cub{p}$ and $\cub{q}$ for
$1\leq p<q\leq\infty$:

\begin{itemize}
\item The carrier of $\cub{p}$ is a proper subset of the carrier of
  $\cub{q}$, and the inclusion of $\cub{p}$ into $\cub{q}$ is a short
  map in $\MS$.

\item The carrier of ${\cub{p}}_n$ is a proper subset of the carrier of ${\cub{q}}_n$,
      and the inclusion of ${\cub{p}}_n$ into ${\cub{q}}_n$ is a short map in
      $\KMS$.
    \item As a topological space, ${\cub{p}}_n$ is a closed sub-space
      of ${\cub{q}}_n$ in $\KH$, since the distances $d_p$ and $d_q$
      induce the same topology on $\Real^n$. Thus, $\kcub{p}$ is a
      closed sub-space of $\kcub{q}$ in $\KH$.

  \end{itemize}
  The last point implies that the carrier of $\kcub{p}$ depends on
  $p$, but the topology does not, namely, it is the topology induced by
  that on $\kcub{\infty}$.

\begin{table}[t!]
 Notation:%
  \begin{tikzcd}
    {} \arrow[r, hook] & {}
  \end{tikzcd}(sub-space) and%
  \begin{tikzcd}
    {} \arrow[r, rightarrowtail] & {}
  \end{tikzcd}(sub-object) in $\MS$ (left) and $\Haus$ (right);
  $X^I$ product of $I$ copies of $X$ (in $\MS$ or $\Haus$), the forgetful
  functor from $\MS$ to $\Haus$ preserves only finite products.
  \begin{center}
  \begin{tikzcd}
    \cub{\infty} \cong \Real_1^\omega \arrow[r, hook] & \ssp{\infty}
    & \ssp{m,\infty}\cong\Real^m \arrow[l, hook']\\
    \cub{p} \arrow[u, rightarrowtail] \arrow[r, hook] & \ssp{p}
    \arrow[u, rightarrowtail] & \ssp{m,p} \arrow[l, hook'] \arrow[u, rightarrowtail]
  \end{tikzcd}
  \quad
  \begin{tikzcd}
    \kcub{p} \arrow[r, hook] & \kcub{\infty} \cong \Real_1^\omega
    \arrow[r, hook] & \kssp{\infty}\cong\cl{\Real}^\omega &
    \kssp{m,\infty}\cong\cl{\Real}^m \arrow[l, hook']\\
    & \cub{\infty} \arrow[u, rightarrowtail] \arrow[r, hook]
    &\ssp{\infty} \arrow[u, rightarrowtail]
    &\ssp{m,\infty}\cong\Real^m \arrow[u, hook'] \arrow[l, hook']\\
    & \cub{p} \arrow[uul, rightarrowtail] \arrow[u, rightarrowtail] \arrow[r, hook] &\ssp{p}
    \arrow[u, rightarrowtail] & \ssp{m,p} \arrow[l, hook'] \arrow[u, equal]
  \end{tikzcd}
  \end{center}

  In the following table, for each complete metric space $\State$
  considered in Section~\ref{sec:examples} (first column), we give:
\begin{itemize}
\item the compact metric sub-space $\State_n$ (second column), \ie, its
  $n$th-approximant;
\item the metric sub-space $\State_*$ (third column), \ie, the union of
  its approximants;
\item the Hausdorff space corresponding to $\State$ (fourth column);
\item the compact Hausdorff sub-space corresponding to $\State_n$ (fifth column);
\item the compact Hausdorff space $\cl{\State}$ given by our construction (sixth column);
\item the property of the map $\iota :\State\to \cl{\State}$ in
  $\Haus$ (seventh column);
\item the section where the example is explained in details (eighth
  column).
\end{itemize}
\[\arrayoptions{1ex}{1.3}
\begin{array}{|*{7}{c|}c|} \hline
\multicolumn{3}{|c|}{\MS}&\multicolumn{4}{c|}{\Haus}&\mbox{see}
\\\hline
\State&\State_n&\State_*&
                          \State&\State_n&\cl{\State}& \iota:\State\to
                                                       \cl{\State} &\mbox{Section}
\\\hline
\multicolumn{8}{c}{\mbox{finite-dimensional Banach spaces}}
\\\hline
\Real&\Real_n&\Real&
\Real&\Real_n&\cl{\Real}&\mbox{sub-space}&
\ref{subsec:examples_real_line}
\\\hline
  \begin{tikzcd}
    \ssp{m,\infty}\arrow[r, equal] & \Real^m
  \end{tikzcd}
  &\Real_n^m&\Real^m&
\multirow{2}{*}{$\Real^m$}&\multirow{2}{*}{$\Real_n^m$}&
\multirow{2}{*}{$\cl{\Real}^m$}&\multirow{2}{*}{\mbox{sub-space}}&
                                                                   \ref{subsec:examples_ell_m_infty}\\
  \begin{tikzcd}
    \ssp{m,p}\arrow[r, rightarrowtail] & \Real^m
  \end{tikzcd}
&(\absn{\Real_n^m},d_p)&\ssp{m,p}&&&&&
\ref{subsec:examples_ell_m_p}
\\\hline
\multicolumn{8}{c}{\mbox{infinite-dimensional Banach spaces}}
\\\hline
\ssp{\infty}&\Real_n^n&\ssp{*,\infty}&\ssp{\infty}&
\multirow{2}{*}{$\Real_n^n$}&\multirow{2}{*}{$\cl{\Real}^\omega$}&
\multirow{2}{*}{\mbox{sub-object}}&
                                    \ref{subsec:examples_ell_infty}\\
  \begin{tikzcd}
    \ssp{p}\arrow[r, rightarrowtail] & \ssp{\infty}
  \end{tikzcd}
&(\absn{\Real_n^n},d_p)&\ssp{*,p}&\ssp{p}&&&&
\ref{subsec:examples_ell_p}
\\\hline
\multicolumn{8}{c}{\mbox{closed bounded convex subsets of infinite-dimensional Banach spaces}}
  \\\hline
  \begin{tikzcd}
    \cub{\infty}\arrow[r, equal] & \Real_1^\omega
  \end{tikzcd}
&\cub{n,\infty}=\Real_1^n&\cub{*,\infty}&\cub{\infty}&
\Real_1^n&\Real_1^\omega&\multirow{2}{*}{\mbox{sub-object}}&
                                                             \ref{subsec:examples_ell_infty_unit_ball}\\
  \begin{tikzcd}
    \cub{p}\arrow[r, rightarrowtail] & \Real_1^\omega
  \end{tikzcd}
&\cub{n,p}&\cub{*,p}&\cub{p}&\cub{n,p}&\kcub{p}&&
\ref{subsec:examples_ell_p_unit_ball}
\\\hline
\end{array}\]
\caption{Summary of Examples of Section~\ref{sec:examples}.}
\label{table:summary_of_examples}
\end{table}

\section{Precision}\label{sec:precision}

Table~\ref{table:summary_of_examples} gives a summary of the examples in
Section~\ref{sec:examples}. We observe the following:

\begin{enumerate}
\item In the finite-dimensional cases, $\State$ is a dense sub-space
  of $\cl{\State}$ in $\Haus$, and $\cl{\State}$ is a compactification
  of $\State$. Therefore, every closed subset $C$ of $\State$ is the
  intersection $C' \bigcap \State$ fore some closed subset $C'$ of
  $\cl{\State}$.  Since the distances $d_p$ and $d_\infty$ induce the
  same topology on $\Real^m$---the carrier set of both $\ssp{m,p}$ and
  $\ssp{m,\infty}$---we have $\cL(\ssp{m,p})=\cL(\ssp{m,\infty})$ for
  each $p:[1, \infty]$.  Furthermore, $\cl{\State}$ does not depend on
  $p$. Hence, it suffices to consider the cases
  $\State=\ssp{m,\infty}$. The map $\iota: \ssp{m,\infty} \to
  \kssp{m,\infty}$ is a sub-space inclusion and $\iota^*\circ \iota_*
  = \id_{\cL(\ssp{m,\infty})}$.

\item In the infinite-dimensional cases---as we will
  demonstrate---$\State$ is not a sub-space of $\cl{\State}$.  More
  precisely, $\iota: \State \to \cl{\State}$ is mono and epi, thus the
  image of $\iota$ is dense in $\cl{\State}$.  However, $\iota$ is not
  a sub-space inclusion, thus there are closed subsets $C$ of $\State$
  which cannot be written as $C' \bigcap \State$ for some closed subset
  $C'$ of $\cl{\State}$, and $\iota^*\circ \iota_*$ is not the
  identity on $\cL(\State)$.
\end{enumerate}

In what follows, we focus mainly on the case of the unit ball
$\cub{p}$. As discussed in
Sections~\ref{subsec:examples_ell_infty_unit_ball}
and~\ref{subsec:examples_ell_p_unit_ball}, without loss of generality,
we assume that the bijective map $\iota:\cub{p}\to \kcub{p}$ is an
identity. As a result, by going from $\cub{p}$ to $\kcub{p}$, the
carrier set does not change, and we have to compare only the
topologies, or equivalently $\cL(\kcub{p})\subset\cL(\cub{p})$.  Since
the left adjoint $\iota^*$ is the inclusion map of $\cL(\kcub{p})$
into $\cL(\cub{p})$, we have $\iota^*(\iota_*(C))=\iota_*(C)$.
Thus, the loss of precision is measured by how bigger is
$\iota_*(C)$ in comparison to the closed subset $C$ in $\cL(\cub{p})$.

\begin{remark}
  \label{remark:Unit_Ball_WLOG}
  In applications, it is reasonable to restrict to closed bounded
  subsets of Banach spaces, \ie, closed subsets included in a ball of
  finite radius.  The claims in this section are true for closed balls
  in $\ssp{p}$ with center $x$ and radius $r>0$, but we state them for
  the paradigmatic case $x=0$ and $r=1$, to avoid extra parameters in
  notation.
  Note that, compact subsets of a Banach space are always closed and
  bounded, and in the finite-dimensional case the converse also holds.
\end{remark}

To start, we present a positive result where there is no loss of precision.
\begin{proposition}[Compact sets]
  \label{prop:compact}
  If%
  \begin{tikzcd}
    \State_1\arrow[r, rightarrowtail] & \State_2
  \end{tikzcd}in $\Haus$, then compact subsets of
  $\State_1$ are compact in $\State_2$.
\end{proposition}
\begin{proof}
  Write $\tau_i$ for the topology on $\State_i$ and $\iota$ for the mono%
  \begin{tikzcd}
    \State_1\arrow[r, rightarrowtail] & \State_2%
  \end{tikzcd}which we can assume to be an inclusion between the
  carriers. Let $\iota^*(\tau_2)$ be the topology on the carrier of
  $\State_1$ induced by $\tau_2$. Since $\iota$ is continuous, we have
  $\iota^*(\tau_2)\subseteq\tau_1$.  Therefore, if $K$ is
  $\tau_1$-compact and $U$ is an open cover of $K$ in
  $\iota^*(\tau_2)$, then $U$ is also an open cover of $K$ in
  $\tau_1$. Hence, it has a finite sub-cover $U_0$.
\end{proof}

\begin{corollary}
  Assume that $\State:\Haus$, and $\iota:\State\to \ol{\State}$ (in
  $\Haus$) is as in Theorem~\ref{thm:idem}. If $C:\cL(\State)$ is
  compact, then $C:\cL(\ol{\State})$, and therefore $\iota_*(C)=C$.
\end{corollary}
\begin{proof}
  As%
\begin{tikzcd}
  \State \arrow[r, rightarrowtail] & \ol{\State},
\end{tikzcd}by Proposition~\ref{prop:compact}, every compact
$C:\cL(\State)$ is also compact in $\ol{\State}$. The result now follows
from the fact that in Hausdorff spaces, compact subsets are closed.
\end{proof}

Compact subsets of infinite-dimensional Banach spaces are not so
relevant in applications (\eg, they always have empty interior). For
$1 < p < \infty$, however, we have a characterization of the closed
bounded subsets of $\ssp{p}$ for which there is no loss of precision,
{\ie}, $C=\iota^*(\iota_*(C))$.

\begin{theorem}[\textbf{No loss of Precision}]
  \label{thm:preservation_of_precision} For any $1 < p < \infty$ and
  $C:\cL(\ssp{p})$, the following are equivalent:
  \begin{enumerate}
  \item\label{item:thm:no_loss}
  $C$ is bounded and $C=\iota^*(\iota_*(C))$, {\ie}, there is no loss of
  precision over $C$.
\item\label{item:thm:inter_union_balls}
  $C$ is a non-empty intersection of finite unions of closed balls in $\ssp{p}$.
  \item\label{item:thm:inter_union_convex} $C$ is a non-empty intersection of
  finite unions of bounded-closed-convex subsets of
  $\ssp{p}$.
  \end{enumerate}
\end{theorem}
\begin{proof}
  See the proof \vpageref{proof:precision_Theorem}.
\end{proof}
To prove Theorem~\ref{thm:preservation_of_precision}, we relate the
topology on $\kcub{p}$ (indeed on any closed ball in $\ssp{p}$) to the
weak-* topology, one of the fundamental topologies studied in
functional analysis.

\subsection{Weak-* topology on  $\absn{\cub{p}}$ for $1 \leq p \leq \infty$}
\label{subsec:weak_star_topology}

Let us first present a quick reminder of weak and weak-* topologies
for the case of normed vector spaces over the field of real
numbers. The reader may refer to any standard book on functional
analysis, {\eg},
\cite{Rudin:Functional_Analysis:Book:1991,Conway:Fun_Analysis:2ed:1990},
for the more general treatment of these topologies.

\begin{definition}[dual]
  Given a normed vector space $X:\NVS$ with norm $\norm{.}_X$,
  its continuous \textbf{dual} $X'$ is the Banach space of linear continuous
  functions from $X$ to $\Real$ with norm
  $\norm{f}_{X'} \defeq \sup \setbarNormal{\absn{f(x)}}{x:X \wedge \norm{x}_{X} \leq 1}$.
\end{definition}

\begin{proposition}
\label{prop:Cauchy_Compl_Dual}
  If $Y:\BS$ is the
  Cauchy completion of $X:\NVS$, then $Y'$ and $X'$ are isomorphic.
\end{proposition}

\begin{proof}
  See, {\eg},~\cite[Page~270]{Narici_Beckenstein:TVS:Book:2011}.
\end{proof}

\begin{definition}[reflexive, weak, weak-*]
  \label{def:weak_star_reflexive}
  For $X:\NVS$, the map $\eta_X:X\to X''$ (a linear isometry) is
  defined as $\eta_X(x)(f)\defeq f(x)$ for $x:X$ and $f:X'$.  When the
  map $\eta_X$ is an iso, $X$ is called \textbf{reflexive}.
\begin{enumerate}
  \item The \textbf{weak} topology $W_X$ on $X$ is the coarsest topology
    making each $f \in X'$ continuous.
  \item The \textbf{weak-*} topology $W_X^*$ on $X'$ is the coarsest
    topology on $X'$ making $\eta_X(x)$ continuous for each $x:X$.
\end{enumerate}
\end{definition}

\begin{proposition}
  Let $p'$ denote the \textbf{conjugate} of $p:[1,\infty]$, \ie,
  the unique $q:[1,\infty]$ such that $1/p+1/q=1$.
\begin{enumerate}

\item \label{item:Cauchy_Compl_p} For every $p:[1,\infty)$, the Cauchy
  completion of $\ssp{*,p}$ is the Banach space $\ssp{p}$.

\item \label{item:Cauchy_Compl_Inf}
The Cauchy completion of $\ssp{*,\infty}$ is the Banach sub-space
$c_0$ of $\ssp{\infty}$.

\item \label{item:linear_iso} For every $p:[1,\infty]$, the map
  $\xi:\ssp{p'}\to (\ssp{*,p})'$, given by
  $\xi(x')(x)\defeq\sum_{i:\omega} x'_i*x_i$, is an isomorphism.

\end{enumerate}

\end{proposition}

\begin{proof}

  Proofs of~\ref{item:Cauchy_Compl_p} and \ref{item:Cauchy_Compl_Inf}
  are straightforward. To prove~\ref{item:linear_iso}, by
  Proposition~\ref{prop:Cauchy_Compl_Dual}, and
  by~\ref{item:Cauchy_Compl_p} and \ref{item:Cauchy_Compl_Inf}, we may
  regard $\xi$ as a function from $\ssp{p'}$ to $(\ssp{p})'$, when
  $p:[1,\infty)$, or to $(c_0)'$, when $p = \infty$. The proof that
  $\xi$ is an isomorphism may now be found in:

  \begin{itemize}
  \item \cite[Appendix B]{Conway:Fun_Analysis:2ed:1990}, for
    $p:[1,\infty)$.

  \item \cite[Page 50]{Albiac_Kalton:Banach_Theory:Book:2006}, for
    $p=\infty$.
  \end{itemize}

\end{proof}

The duals and double duals of relevance in this section are summarized
in Table~\ref{tab:duals_double_duals}.
\begin{table}[t]
  \centering
  \arrayoptions{1ex}{1.3}
  \begin{tabular}{|c|c|c|c|}\hline
    $\boldsymbol{X}$ & $\boldsymbol{X'}$ & $\boldsymbol{X''}$ & \textbf{Note} \\ \hline \hline
    $c_0$ & $\ell_1$ & $\ell_\infty$ & \\ \hline
    $\ell_p$ & $\ell_{p'}$ & $\ell_p$ & $1 < p < \infty$ \\ \hline
  \end{tabular}
  \caption{Some duals and double duals (up to iso).}
  \label{tab:duals_double_duals}
\end{table}

\begin{definition}[Topologies $\tau_p, \tau_p^*, {\cl{\tau}}_p$]
    \label{def:tau_p_notations}
We define the following topologies on the carrier of $\ssp{p}$:
  \begin{enumerate}

  \item $\tau_p$ is the original (or, norm) topology on $\ssp{p}$, \ie, the
  topology induced by the norm $\norm{.}_p$.

\item $\tau_p^*$ denotes the weak-* topology on $\ssp{p}$ as the
  continuous dual of $\ssp{*,p'}$.

  \item ${\cl{\tau}}_p$ is the topology on $\ssp{p}$ as a subset of the
  compact Hausdorff space $\cl{\Real}^\omega$.
  \end{enumerate}
We use the same notation for the topologies when restricted to a subset of
$\ssp{p}$, such as $\cub{p}$.
\end{definition}
According to Table~\ref{table:summary_of_examples}, by using the notation of
Definition~\ref{def:tau_p_notations}, we have that $\cl{\State} =
(\absn{\cub{p}}, {\bar{\tau}}_p)$ when $\State = (\absn{\cub{p}},
\tau_p)$. Therefore, we obtain:
\begin{proposition}
  \label{prop:bar_tau_p_closure}
    For every $C:\cL(\ssp{p})$, the set $\iota^*(\iota_*(C))$ is the
    ${\cl{\tau}}_p$ closure of $C$.
\end{proposition}

In Theorem~\ref{thm:weak_star} we show that the topological spaces
$(\absn{\cub{p}}, {\bar{\tau}}_p)$ and $(\absn{\cub{p}}, \tau_p^*)$
coincide for any $1 \leq p \leq \infty$.  First, we recall the
lemma on the `rigidity' of compact Hausdorff
topologies in~\cite[Section~3.8]{Rudin:Functional_Analysis:Book:1991}.

\begin{lemma}
    \label{lemma:rigidity_compact_Hausdorff}
    If $\tau_1 \subseteq \tau_2$ are topologies on a set $X$, with
    $\tau_1$ Hausdorff and $\tau_2$ compact, then $\tau_1 = \tau_2$
\end{lemma}
\begin{proof}
    Let $F \subseteq X$ be $\tau_2$-closed. Since $X$ is
    $\tau_2$-compact, so is $F$. Since $\tau_1 \subseteq \tau_2$, it
    follows that $F$ is $\tau_1$-compact. Since $\tau_1$ is a
    Hausdorff topology, it follows that $F$ is $\tau_1$-closed.
\end{proof}

\begin{theorem}
    \label{thm:weak_star}
    For each $1 \leq p \leq \infty$, $\kcub{p}$ is (isomorphic to) the
    topological space $(\absn{\cub{p}}, \tau_p^*)$.
\end{theorem}

\begin{proof}
    The topology ${\cl{\tau}}_p$ is Hausdorff. On the other hand, by
    the Banach-Alaoglu theorem (see,
    {\eg},~\cite[Section~3.8]{Rudin:Functional_Analysis:Book:1991})
    the closed unit ball $\absn{\cub{p}}$ is weak-* compact, \ie,
    $\tau_p^*$-compact. Therefore, by
    Lemma~\ref{lemma:rigidity_compact_Hausdorff}, it suffices to prove
    that ${\cl{\tau}}_p \subseteq \tau_p^*$.

    For $i:\omega$, consider the projections
    $\pi_i : \ssp{p} \to \Real$ defined by

  \begin{equation}
    \label{eq:projections_pi}
  \forall x: \ssp{p}: \pi_i(x) \defeq x_i.
\end{equation}
\noindent
The set
${\cal Y} = \setbarNormal{ \absn{\cub{p}} \bigcap \pi_i^{-1}(O) }{i:\omega,
  O \subseteq \Real \text{ Euclidean open}}$ is a sub-base for
${\cl{\tau}}_p$. For each $i:\omega$, consider the sequence $e_i$
defined by

\begin{equation}
  \label{eq:e_i}
  \forall n: \omega. \quad e_i(n) \defeq
  \left\{
    \begin{array}{ll}
      0,& \text{ if } i \neq n,\\
      1, & \text{ if } i = n.\\
    \end{array}
  \right.
\end{equation}
  \noindent
  For all $i:\omega$, the sequence $e_i$ is in all the relevant
  pre-duals as specified in Definition~\ref{def:tau_p_notations}
  above. Furthermore

  \begin{equation*}
    \forall i:\omega. \forall x: \ssp{p}: \quad \pi_i(x) = x(e_i).
  \end{equation*}
  Thus, each $\pi_i$ is weak-* continuous and
  ${\cal Y} \subseteq \tau_P^*$, which entails that
  ${\cl{\tau}}_p \subseteq \tau_P^*$.
  \end{proof}

  \begin{remark} As pointed out
    in Remark~\ref{remark:Unit_Ball_WLOG}, although we present results
    for closed unit balls, they hold for arbitrary closed balls.  In
    particular, Theorem~\ref{thm:weak_star} holds for closed balls in
    $\ssp{p}$, because they are all $\tau_p^*$-compact.
  \end{remark}

  \begin{corollary}
    \label{cor:weak_star_closed}
    For each $1 \leq p \leq \infty$ and $C:\cL(\cub{p})$, $\iota_*(C) = C$ iff $C$ is a $\tau_p^*$-closed subset of $\absn{\cub{p}}$.
  \end{corollary}

  \begin{proof}
    Follows from Proposition~\ref{prop:bar_tau_p_closure} and Theorem~\ref{thm:weak_star}.
  \end{proof}

  \begin{corollary}
    \label{corollary:reflexive_weakly_closed}
    For each $1 < p < \infty$ and $C:\cL(\cub{p})$,
    $\iota_*(C) = C$ iff $C$ is a weakly-closed subset of
    $\absn{\cub{p}}$.
  \end{corollary}

  \begin{proof}
    Since the space $\ssp{p}$ for $1 < p < \infty$ is reflexive,
    {\ie}, the weak and weak-* topologies on $\ssp{p}$ coincide. The
    result follows from Corollary~\ref{cor:weak_star_closed}.
  \end{proof}

  We have established all the preliminaries for presenting the proof
  of Theorem~\ref{thm:preservation_of_precision}:

  \begin{proof}\textbf{(Theorem~\ref{thm:preservation_of_precision})}

    \label{proof:precision_Theorem}

    \begin{description}
    \item[(\ref{item:thm:no_loss}) $\Rightarrow$
      (\ref{item:thm:inter_union_balls})] As $C$ is assumed to be
      bounded, then it must be a subset of a closed ball $B$ of
      finite radius. If $\iota^*(\iota_*(C)) = C$, then, by
      Corollary~\ref{corollary:reflexive_weakly_closed}, $C$ must be
      weakly closed. Thus, $C$ is a weakly closed subset of (the
      weakly compact set) $B$. As a result, it is weakly compact.

      According to~\cite{Corson_Lindenstrauss:weakly_compact:1966}, a
      subset of a separable reflexive space is weakly compact if and
      only if it is the non-empty intersection of finite unions of
      closed balls. Each space $\ell_p$ for $1 < p < \infty$ is
      separable and reflexive. Hence, the result follows.

    \item[(\ref{item:thm:inter_union_balls}) $\Rightarrow$
      (\ref{item:thm:inter_union_convex})] This is straightforward as
      every closed ball is a bounded-closed-convex subset.

    \item[(\ref{item:thm:inter_union_convex}) $\Rightarrow$
      (\ref{item:thm:no_loss})] Assume that $C$ is a non-empty
      intersection of finite unions of bounded-closed-convex subsets
      of $\ssp{p}$. To be precise,
      $C = \bigcap_{i:I} \bigcup_{j:k_i} C_{i,j}$ with $k:\omega^I$
      and $C_{i,j}$ bounded-closed-convex subset of $\ssp{p}$.

      As a consequence of Hahn-Banach separation theorem, every closed
      and convex subset of a Banach space is weakly closed (see,
      {\eg},~\cite[Theorem~3.12]{Rudin:Functional_Analysis:Book:1991}). This
      entails that each $C_{i,j}$ is weakly closed. As the set of
      closed subsets (under any topology) are closed under finite
      unions and arbitrary intersections, then $C$ itself is also
      weakly closed. Clearly, $C$ is also bounded. The result now
      follows from Corollary~\ref{corollary:reflexive_weakly_closed}.
    \end{description}
  \end{proof}

  Item~(\ref{item:thm:inter_union_convex}) of
  Theorem~\ref{thm:preservation_of_precision} provides examples of
  practical importance where no loss of precision is incurred.

\begin{example}[Sequence intervals]
  For each pair $s,t:\ssp{p}$, we define the sequence interval $[s,t]$
  by:
  \begin{equation*}
  [s,t] \defeq \setbarNormal{u:\ssp{p}}{\forall i:\omega. \ s_i \leq
    u_i \leq t_i}.
\end{equation*}
\noindent
In general, sequence intervals are not norm-compact in $\ssp{p}$. They
are, however, bounded, \emph{norm-closed, and convex}. Hence, when
$1 < p < \infty$, there is no loss of precision over sequence
intervals, or indeed, over any subset $C$ of $\ell_p$ which may be
written as an intersection of finite unions of sequence intervals.
\end{example}

\subsubsection{Metrizability of the weak-* topology over closed balls}
\label{subsubsec:metrizability}

Consider the map
$d_* : \ssp{\infty} \times \ssp{\infty} \to \Real$ defined as follows:
\begin{equation}
  \label{eq:d_star_metric}
  \forall x, y : \ssp{\infty}. \ d_*( x, y) \defeq \sum_{n:\omega} \frac{d(x_n,y_n)}{2^{n+1}}.
\end{equation}
We prove that $d_*$ is a metric on the carrier of $\ssp{\infty}$,
which induces the topology on $\kssp{\infty}$.

\begin{proposition}
  \label{prop:ball_weak_start_d_star_infty}
  For any closed ball $B$ of finite radius in $\ssp{\infty}$, the weak-*
  topology on $B$ is induced by $d_*$.
\end{proposition}
\begin{proof}
  According to~\cite[Section~3.8(c),
  page~63]{Rudin:Functional_Analysis:Book:1991}, if $X$ is a compact
  topological space and if some uniformly bounded sequence
  $(f_n \mid {n:\omega})$ of continuous real-valued functions separates
  points on $X$, then $X$ is metrizable, with the metric
  $d(x,y) = \sum_{n:\omega} 2^{-n} \absn{f_n(x) - f_n(y)}$.

  By the Banach-Alaoglu theorem, the ball $B$ is weak-* compact. The
  countable set $\setbarNormal{\pi_i}{i:\omega}$ of projections
  from~\eqref{eq:projections_pi} is separating on $B$, in the
  sense that:
\begin{equation*}
  \forall x,y : B. \ x \neq y \Rightarrow \exists i:\omega. \pi_i(x)
  \neq \pi_i(y).
\end{equation*}
\noindent
Furthermore, each $\pi_i$ is weak-* continuous. Thus, it
suffices to take $f_n = \pi_n/2$, and the claim follows
from~\cite[Section~3.8(c),
page~63]{Rudin:Functional_Analysis:Book:1991}.
\end{proof}

The metric $d_*$ on the carrier of $\ssp{\infty}$ satisfies the
following property
\begin{equation}
  \label{eq:d_star_leq_d_p}
  \forall p:[1,\infty].\,
  \forall x, y : \ssp{p}. \ d_*( x, y) = \sum_{n:\omega}
  \frac{d(x_n,y_n)}{2^{n+1}} \leq \sum_{n:\omega}
  \frac{d_\infty(x,y)}{2^{n+1}} = d_\infty(x,y) \leq d_p(x,y).
\end{equation}

\begin{corollary}
  \label{cor:ball_weak_start_d_star_p}
  For each $p:[1,\infty]$, the weak-* topology on a closed ball of
  finite radius in $\ssp{p}$ is induced by $d_*$.
\end{corollary}

\begin{proof}
  Let $B$ be the closed ball in $\ssp{p}$ centered at $x_0$ with
  radius $r>0$. Let $B'$ denote the closed ball in $\ssp{\infty}$
  centered at $x_0$ with radius $r>0$. The weak-* topology on $B$ is
  the relative topology induced from the weak-* topology on $B'$. The
  claim now follows from
  Proposition~\ref{prop:ball_weak_start_d_star_infty}.
\end{proof}

By going from $\cub{p}$ to $\kcub{p}$, robustness with respect to
the metric $d_p$ is replaced by robustness with respect to the metric
$d_*$. Indeed, inequality~\eqref{eq:d_star_leq_d_p} shows that for any
subset $S$ of $\absn{\cub{p}}$ its $\delta$-neighborhood $B(S,\delta)$
under $d_p$ is included its $\delta$-neighborhood under $d_*$.

  \begin{proposition}\label{prop:eC_d}
    For every $C:\cL(\cub{p})$:
    \begin{enumerate}
    \item \label{prop:item:eC_clos_d_star} The set $\iota_*(C)$ is the closure of $C$ under the $d_*$
      metric.
    \item For every $x:\cub{p}$, $x\in \iota_*(C)$ if and only if:
  \begin{equation}\label{eq:eC_d}
  \forall n:\omega.\forall \delta>0.  \exists y:C.\forall
  i:n. \quad d(x_i,y_i)<\delta .
\end{equation}
\end{enumerate}
\end{proposition}

\begin{proof} {\ }
  \begin{enumerate}
  \item Follows from Corollary~\ref{cor:ball_weak_start_d_star_p}.
  \item From \ref{prop:item:eC_clos_d_star}, we deduce that
    $x\in \iota_*(C)$ if and only if:

    \begin{equation}
      \label{eq:eC_d_star}
      \forall \epsilon > 0. \exists y:C. \quad d_*( x, y) < \epsilon.
    \end{equation}
Thus, it suffices to prove that \eqref{eq:eC_d} and
\eqref{eq:eC_d_star} are equivalent.

\begin{description}

\item[\eqref{eq:eC_d} $\Rightarrow$ \eqref{eq:eC_d_star}:] For any
  given $\epsilon > 0$, choose $n$ large enough such that
  $\epsilon 2^{n-2} > 1$ and $\delta = \epsilon /
  2$. By~\eqref{eq:eC_d}, there exists a $y:C$ such that
  $\forall i:n. \ d(x_i,y_i)<\delta$. For this $y$ we have:

  \begin{align*}
    d_*(x,y) &= \sum_{i:n} \frac{d(x_i, y_i)}{2^{i+1}} +  \sum_{i \geq
               n} \frac{d(x_i, y_i)}{2^{i+1}} \\
    &\leq \sum_{i:n} \frac{\delta}{2^{i+1}} +  \sum_{i \geq
      n} \frac{2}{2^{i+1}} \\
             &\leq \delta  +  \frac{2}{2^n} \leq \frac{\epsilon}{2}  +  \frac{\epsilon}{2} = \epsilon.
  \end{align*}

\item[\eqref{eq:eC_d_star} $\Rightarrow$ \eqref{eq:eC_d}:] For any
  given $n:\omega$ and $\delta > 0$, choose
  $\epsilon = \delta / 2^{n+1}$. By~\eqref{eq:eC_d_star}, there exists
  a $y:C$ such that $d_*(x,y) < \epsilon$, which implies that:

  \begin{equation}
    \label{eq:d_star_epsilon}
    \sum_{i:\omega} \frac{d(x_i, y_i)}{2^{i+1}} < \epsilon = \frac{\delta}{2^{n+1}}.
  \end{equation}
  From~\eqref{eq:d_star_epsilon}, we obtain $\forall i:\omega. d(x_i,
  y_i) < \delta*2^{i-n}$. In particular, for every $i:n$, we have $d(x_i,
  y_i) < \delta$.
\end{description}
  \end{enumerate}
\end{proof}

\subsection{Loss of precision}
\label{subsec:loss_of_precision}

In this subsection, we discuss loss of precision through a few
examples.

\begin{example}[bounded, closed, non-convex, discrete]
  \label{example:discrete_e_i_s} Take the sequences
  $(e_i \mid i:\omega)$ as defined in~\eqref{eq:e_i}, {\ie}:
\begin{equation*}
  \forall n: \omega. \quad e_i(n) \defeq
  \left\{
    \begin{array}{ll}
      0,& \text{ if } i \neq n,\\
      1, & \text{ if } i = n,\\
    \end{array}
  \right.
\end{equation*}
\noindent
and consider the set $C \defeq \setbarNormal{e_i}{i:\omega}$, which is
a discrete, bounded, and closed subset of $\cub{p}$ for
$p:[1,\infty]$. Using the metric $d_*$ of~\eqref{eq:d_star_metric}, we
show that $\iota_*(C) = C \bigcup \set{0}$.

  It is indeed clear that $0$ is in the weak-* closure of $C$ as:

  \begin{equation*}
    \lim_{i \to \infty} d_*(e_i, 0) = \lim_{i \to \infty} 2^{-(i+2)} = 0.
  \end{equation*}
  Furthermore,
  $\forall i,j:\omega. \ d_*(e_i,e_j) \leq 2^{-(i+2)} +
  2^{-(j+2)}$. This entails that
  $\lim_{i,j \to \infty} d_*(e_i, e_j) = 0$. Hence, the sequence
  $(e_i \mid {i:\omega})$ is a Cauchy sequence under $d_*$ and it can have
  only one limit point, {\ie}, zero.

\end{example}

\begin{example}[bounded, closed, non-convex, connected]
  \label{example:unit_sphere}
  Consider the unit sphere
  $S_p \defeq \setbarNormal{s:\cub{p}}{ \norm{s}_p = 1}$, which is
  norm-closed and non-convex. The weak-* closure of the unit sphere in
  $\absn{\cub{p}}$ is the entire unit ball $\absn{\cub{p}}$. In this
  case, the loss of precision is quite noticeable.
\end{example}

In Theorem~\ref{thm:preservation_of_precision}, the assumption
$p \not \in \set{ 1 , \infty}$ is crucial, as demonstrated in
Example~\ref{example:counter_p_infinity} and
Proposition~\ref{prop:kernel_ell_1}.

\begin{example}[bounded, closed, convex, $p=\infty$]
  \label{example:counter_p_infinity}
  Consider the set $C \defeq c_0 \bigcap \cub{\infty}$, which is a
  bounded, closed, and convex subset of $\cub{\infty}$. In
  Section~\ref{subsec:examples_ell_infty_unit_ball} we proved that
  $\iota_*(C) = \cub{\infty}$.
\end{example}

Theorem~\ref{thm:preservation_of_precision} relies on the fact that
norm-closed and convex subsets of reflexive Banach spaces are weakly
closed. On the other hand, norm-closed and convex subsets of
non-reflexive Banach spaces are not, in general, weak-* closed. Recall
that the map $\eta_X$ in Definition~\ref{def:weak_star_reflexive}
is a linear isometry which embeds $X$ into $X''$. When $X$ is not
reflexive, we have $X'' \setminus X \neq \emptyset$.

\begin{proposition}
  \label{prop:non-reflexive-kernel} Assume that $X$ is a non-reflexive
  Banach space. Then, for any element $\theta \in X'' \setminus X$,
  the kernel $\theta^{-1}(0)$ of $\theta$ is a closed linear
  sub-space of $X'$ which is not weak-* closed.
\end{proposition}

\begin{proof}
  This is a consequence of~\cite[Theorem~3.1,
  page~108]{Conway:Fun_Analysis:2ed:1990} and~\cite[Theorem~3.10,
  page~64]{Rudin:Functional_Analysis:Book:1991}.
\end{proof}

Using the following result, for any given non-reflexive Banach space,
one may construct many examples of sets for which there is a loss of
precision, provided the space is the dual of another Banach space,
{\ie}, has a pre-dual. This rules out spaces such as $c_0$ which have
no pre-duals~\cite[Theorem~6.3.7]{Albiac_Kalton:Banach_Theory:Book:2006}.

\begin{corollary}[bounded, closed, convex, non-reflexive with pre-dual]
  \label{cor:non_reflex_kernel_ball}
  Consider a non-reflexive Banach space $X$ and let $B$ denote the
  closed unit ball of $X'$. For any $\theta \in X'' \setminus X$,
  define $C \defeq B \bigcap \theta^{-1}(0)$. Then, the set $C$ is closed and
  convex---hence, weakly closed---but not weak-* closed.
\end{corollary}

\begin{proof}
  As a consequence of Proposition~\ref{prop:non-reflexive-kernel}, the
  set $C$ is closed and convex, hence, weakly
  closed. By~\cite[Corollary~12.6,
  page~160]{Conway:Fun_Analysis:2ed:1990}, however, $C$ cannot be
  weak-* closed.
\end{proof}

In general, an exact description of the weak-* closures of sets $C$
obtained in Corollary~\ref{cor:non_reflex_kernel_ball} is not
known. There are, however, inner and outer approximations available in
the literature. For instance, according
to~\cite[Proposition~2]{Jameson:weak_star_closure_unit_ball:1982}, the
weak-* closure of $C$ must contain a closed ball centered at the
origin. Proposition~\ref{prop:kernel_ell_1} gives an instance of this
result, which can be proved directly.

\begin{proposition}[bounded, closed, convex, $p=1$]
  \label{prop:kernel_ell_1}
  Take the set $C \defeq \setbarNormal{x:\cub{1}}{\sum_{n:\omega} x_n
    = 0}$ and the closed ball $D$ in $\ell_1$ centered at $0$
  with radius $1/2$. Furthermore,
  let $\cl{C}$ denote the
  $\tau_1^*$-closure of $C$ in $\cub{1}$. Then:
  \begin{enumerate}
  \item \label{prop:item:C_subset_cub1} $C$ is a closed and
    convex---hence, also weakly closed---subset of $\cub{1}$. But, $C$
    is not weak-* closed.

  \item \label{prop:item:half_ball_subset}
    $D \setminus C \neq \emptyset$.
  \item \label{prop:item:C_subset_unit_ball} $D \subset \cl{C} \subset \cub{1}$, where both
    inclusions are strict.
  \end{enumerate}

\end{proposition}

\begin{proof}
  Take $X = c_0$ (which implies that $X' = \ssp{1}$ and
  $X'' = \ssp{\infty}$) and let
  $\theta = [1]^\omega : \ssp{\infty} \setminus c_o$. We have
  $C = \cub{1}  \bigcap \theta^{-1}(0)$.

  \begin{enumerate}
  \item Follows from Corollary~\ref{cor:non_reflex_kernel_ball}.

  \item
    $\left(\frac{1}{2^{n+2}}\mid {n:\omega} \right) \in D \setminus
    C$.

  \item Assume that $x = (x_n \mid {n:\omega}) : D$. For any
    $n:\omega$ and $\delta > 0$, consider
    $\hat{x} = (\hat{x}_i \mid {i:\omega})$ defined as follows:

  \begin{equation*}
    \forall i:\omega.\ \hat{x}_i \defeq
    \left\{
      \arrayoptions{1ex}{1.2}
      \begin{array}{ll}
        x_i, & \text{if } i<n,\\
        -x_{i-n}, & \text{if } n\leq i<2n,\\
        0, & \text{otherwise}.
      \end{array}
    \right.
  \end{equation*}
  We have $\theta(\hat{x}) = \sum_{i:\omega} \hat{x}_i =
  0$. Furthermore, $\norm{\hat{x}}_1 \leq 2 \norm{x}_1 \leq 1$. Hence,
  $\hat{x} \in \theta^{-1}(0) \bigcap \cub{1} = C$. Clearly, for all
  $i:n$, we have $d(x_i, \hat{x}_i) = 0 \leq \delta$. Therefore,
  equation~\eqref{eq:eC_d} of Proposition~\ref{prop:eC_d} holds, and
  we have $D \subseteq \cl{C}$.

  Take the sequence $y = (y_n \mid {n:\omega})$ defined as follows:

  \begin{equation*}
    \forall n:\omega.\ y_n \defeq
    \left\{
      \arrayoptions{1ex}{1.2}
      \begin{array}{ll}
        0.5, & \text{if } n=0,\\
        -0.5, & \text{if } n=1,\\
        0, & \text{otherwise}.
      \end{array}
    \right.
  \end{equation*}
  Then,
  $y \in C \setminus D \subseteq \cl{C} \setminus
  D$. Thus, we have proved that $D \subset \cl{C}$.

  It remains to prove that $\cub{1} \setminus \cl{C} \neq
  \emptyset$. Take the sequence $z = (z_n \mid {n:\omega})$
  defined as follows:

  \begin{equation*}
    \forall n:\omega.\ z_n \defeq
    \left\{
      \arrayoptions{1ex}{1.2}
      \begin{array}{ll}
        1, & \text{if } n=0,\\
        0, & \text{otherwise}.
      \end{array}
    \right.
  \end{equation*}
  Clearly, $z \in \cub{1}$. We use Proposition~\ref{prop:eC_d} to
  prove that $z \not \in \cl{C}$. In~\eqref{eq:eC_d}, choose $n=1$ and
  $\delta = 1/2$. For any given $\hat{z} : C$, we have
  $\sum_{n:\omega} \hat{z}_n = 0$. Thus,
  $\sum_{n=1}^\infty \hat{z}_n = -\hat{z}_0$, which implies that:

  \begin{equation}\label{eq:hat_z_sum_z_0}
  \absn{\sum_{n=1}^\infty \hat{z}_n} = \absn{\hat{z}_0}.
  \end{equation}

  On the other hand, as $\hat{z}:C$, we must have
  $\norm{\hat{z}}_1 \leq 1$. As a result:

  \begin{equation}\label{eq:one_geq_hat_z_0_sum}
    1 \geq\norm{\hat{z}}_1  = \sum_{n:\omega} \absn{\hat{z}_n} \geq
    \absn{\hat{z}_0} + \absn{\sum_{n=1}^\infty \hat{z}_n}.
  \end{equation}
  By combining~\eqref{eq:hat_z_sum_z_0} and
  \eqref{eq:one_geq_hat_z_0_sum}, we obtain
  $\absn{\hat{z}_0} \leq 1/2$, which implies
  $1 - \hat{z}_0 \geq 1/2$, equivalently
  $d(z_0, \hat{z}_0) \geq 1/2 = \delta$.
  \end{enumerate}
\end{proof}


\section{Concluding Remarks}
\label{sec:concluding_remarks}

The results in this paper are part of an overall study of robust maps.
We have chosen the theory of $\omega$-continuous
lattices, within which computability can be studied using the
framework of effectively given domains~\cite{smyth1977effectively},
and robustness can be analyzed using the Robust topology
(Definition~\ref{def-top-S}) over the lattice of closed subsets of the
state space.

In a related work, Edalat~\cite{Edalat95:DT-fractals} has considered
locally compact Hausdorff spaces instead of metric spaces, and has
worked with the domain of compact subsets (ordered by reverse
inclusion) instead of the complete lattice of closed
subsets. Furthermore, he has investigated the relationship between
Scott topology and the upper Vietoris topology, but has not studied
robustness. The Robust topology lies in between Scott and the upper
Vietoris topologies~(Theorem~\ref{thm-top-S}).

The case of compact metric spaces has been studied
in~\cite{Moggi_Farjudian_Duracz_Taha:Reachability_Hybrid:2018}. This
suffices to deal with the input space of typical machine learning
systems, and the state space of common hybrid systems. In this paper,
the focus has been non-compact metric spaces, for which, a novel
approach has been presented based on approximation of the space via a
(growing) sequence of compact metric sub-spaces. Non-compact spaces
are relevant when dealing with perturbations of the \emph{model}
parameters of a system, {\eg}, perturbations of the activation
function(s) of a neural network, or the flow $F$ and jump $G$
relations of a hybrid system $(\State,F,G)$.

We presented a detailed account of some examples, including (closed
bounded subsets of) infinite-dimensional Banach spaces, and analyzed
the important issue of precision, when it is retained, and when
precision is lost. In particular, we have obtained a complete
characterization of the closed subsets of reflexive spaces~$\ssp{p}$
(\ie, those with $1 < p < \infty$), for which there is no loss of
precision~(Theorem~\ref{thm:preservation_of_precision}).

All examples studied in this paper are sequence spaces. As such,
studying other relevant spaces provides an immediate direction for
future work. For instance, let $\Omega \subseteq \Real^n$ be an open
set.  Lebesgue spaces $L^p(\Omega)$ are examples for future work,
which are relevant in the study of partial differential
equations~\cite{Brezis:Functional_Analysis-Book:2011}.

Other cases for future study include infinite-dimensional feature
spaces arising in machine learning and spaces of bounded measures. In
particular, by applying our results to (closed subsets of) probability
measures, we obtain a framework for computation of probability
measures using finitary approximations. It will be interesting to
compare the finitary approximations obtained in this way, with those
obtained by Edalat~\cite{Edalat:Scott_weak_on_top:1997} for
computation of probability measures over separable metric spaces.

\bibliographystyle{plain}
\bibliography{Biblio.bib}

\end{document}